\documentclass[sigplan, nonacm, screen]{acmart}
\AtBeginDocument{%
  }

\usepackage{graphicx}
\usepackage{subcaption}
\usepackage{algorithm}
\usepackage{algpseudocode}
\usepackage{setspace}

\setcopyright{acmlicensed}
\copyrightyear{2027}
\acmYear{2027}
\acmDOI{XXXXXXX.XXXXXXX}
\acmConference[Conference acronym 'XX]{Make sure to enter the correct
  conference title from your rights confirmation email}{June 03--05,
  2018}{Woodstock, NY}
\acmISBN{978-1-4503-XXXX-X/2018/06}

\begin{document}

\title[SpiderLS]{SpiderLS: Leveraging Full ZX Reduction\\for Lattice Surgery Compilation}


\author{Hyungseok Kim}
\orcid{0009-0003-8981-3015}
\affiliation{%
  \institution{Yonsei University}
  \city{Seoul}
  \country{Republic of Korea}}
\email{hyungseok.kim@yonsei.ac.kr}

\author{Changheon Lee}
\orcid{0009-0006-4886-8601}
\affiliation{%
  \institution{Yonsei University}
  \city{Seoul}
  \country{Republic of Korea}}
\email{changheon.lee@yonsei.ac.kr}

\author{Seungjik Kim}
\orcid{0009-0006-2855-3477}
\affiliation{%
  \institution{Yonsei University}
  \city{Seoul}
  \country{Republic of Korea}}
\email{seungjik.kim@yonsei.ac.kr}

\author{Enhyeok Jang}
\orcid{0009-0000-7034-6793}
\affiliation{%
  \institution{Yonsei University}
  \city{Seoul}
  \country{Republic of Korea}}
\email{enhyeok.jang@yonsei.ac.kr}

\author{Youngmin Kim}
\orcid{0009-0002-8346-4830}
\affiliation{%
  \institution{Yonsei University}
  \city{Seoul}
  \country{Republic of Korea}}
\email{youngmin.kim@yonsei.ac.kr}

\author{Seungwoo Choi}
\orcid{0009-0005-2162-8993}
\affiliation{%
  \institution{Yonsei University}
  \city{Seoul}
  \country{Republic of Korea}}
\email{seungwoo.choi@yonsei.ac.kr}

\author{Hanbit Lee}
\orcid{0009-0007-0116-6204}
\affiliation{%
  \institution{Yonsei University}
  \city{Seoul}
  \country{Republic of Korea}}
\email{hanbit.lee@yonsei.ac.kr}

\author{Sungho Pyun}
\orcid{0009-0004-1144-4157}
\affiliation{%
  \institution{Yonsei University}
  \city{Seoul}
  \country{Republic of Korea}}
\email{sungho.pyun@yonsei.ac.kr}

\author{Won Woo Ro}
\correspondingauthor
\orcid{0000-0001-5390-6445}
\affiliation{%
  \institution{Yonsei University}
  \city{Seoul}
  \country{Republic of Korea}}
\email{wro@yonsei.ac.kr}

\renewcommand{\shortauthors}{Kim et al.}

\begin{abstract}

Lattice surgery compilation plays a central role in translating fault-tolerant quantum programs into efficient surface code realizations, where both spatial and temporal resources directly determine the cost of execution.
Recent work has demonstrated the benefits of using ZX-diagrams as an intermediate representation for lattice surgery compilation, enabling semantics-preserving transformations that reduce spacetime cost.
However, existing compilation restricts ZX reduction to preserve diagram structures that can be directly embedded as lattice surgery junctions.
We present SpiderLS, which extends prior approach by leveraging full ZX reduction.
To translate the resulting diagram into executable lattice surgery operations, SpiderLS applies a sequence of compiler passes that derives an execution order, generates target code by grouping compatible interactions into multi-target operations, and lowers the target code to Pauli-product measurements.
The resulting explicit patch and Pauli-boundary requirements guide logical scheduling and structure-aware spacetime routing.
Across representative algorithmic and random workloads, SpiderLS achieves average reductions of 49.2\% in spacetime volume and 99.8\% in compilation time compared with the prior ZX-based compiler.

\end{abstract}




\maketitle

\section{Introduction}

Fault-tolerant quantum computing (FTQC) has attracted growing attention as a path toward executing large-scale quantum programs reliably in the presence of physical errors \cite{gottesman1998theory, campbell2017roads, viszlai2026prophunt}.
Realizing FTQC requires quantum error correction, which encodes logical qubits into larger collections of physical qubits and continuously protects them throughout computation.
Among existing error-correcting codes, the surface code is a leading architecture for FTQC due to its high error tolerance and reliance on local operations \cite{KITAEV20032, PhysRevA.86.032324, google2025quantum, google2023suppressing}.
Lattice surgery provides a practical computational model for the surface code, implementing logical operations by merging and splitting encoded qubit patches through Pauli-product measurements \cite{horsman2012surface, litinski2019game}.
As quantum programs scale, however, translating logical circuits into efficient lattice surgery realizations becomes increasingly important \cite{tan2024sat, watkins2024high, molavi2025dependency}: the compiler must determine how logical operations are represented, scheduled, and geometrically realized while minimizing the resulting spacetime cost \cite{hamada2026efficient, hamada2026bounded, low2026denser}.

Recent work \cite{zhou2026topols} has taken an important step toward efficient lattice surgery compilation by using the ZX-calculus as an intermediate representation \cite{coecke2011interacting, de2020zx}.
ZX-diagrams represent quantum computations as graphs and admit semantics-preserving rewrite rules, allowing a compiler to transform the structure of a computation beyond the constraints of the circuit representation \cite{van2020zx}.
Moreover, the close correspondence between ZX spiders and lattice surgery operations provides a natural path from the transformed representation to a physical spacetime realization \cite{de2020zx, gidney2019flexible, paler2025untangling}.
However, fully exploiting this flexibility is nontrivial: more extensive ZX reduction may produce graph structures that no longer admit a direct geometric interpretation in lattice surgery, making the subsequent lowering to executable lattice surgery operations a key challenge \cite{zhou2026topols}.

\begin{figure*}[ht]
     \centering
     \includegraphics[
        width=0.84\textwidth,
        trim=1.4cm 3.75cm 2.7cm 1.5cm,
        clip
     ]{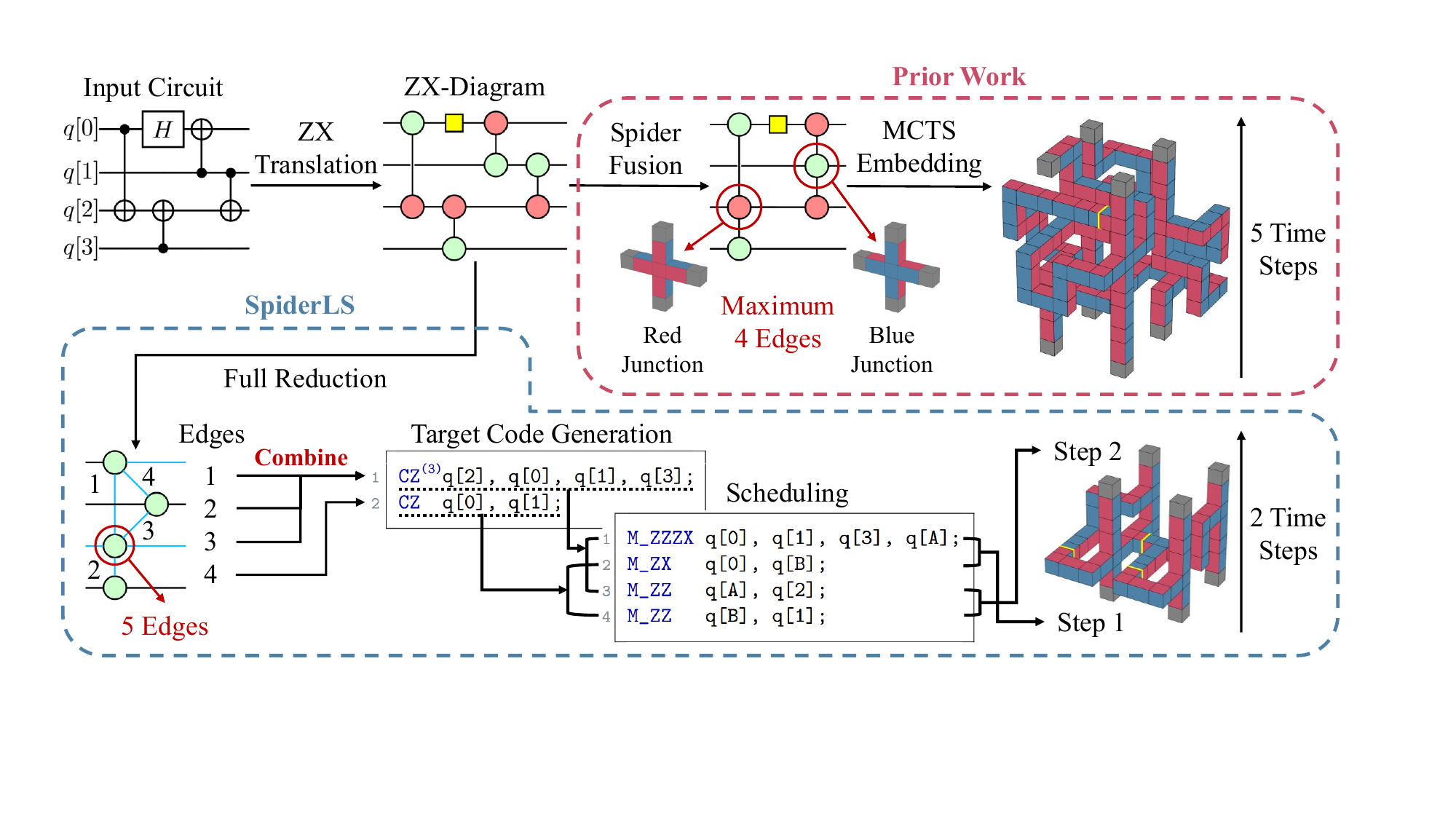}
     \caption{
     Overview of SpiderLS in comparison with prior ZX-based lattice surgery compilation approach \cite{zhou2026topols}.
     Prior work \cite{zhou2026topols} reduces ZX-diagram using only spider fusion and performs layer-wise embedding via MCTS. 
     In contrast, SpiderLS employs full ZX reduction, generates high-arity interactions, and performs layered spacetime routing.
     }
     \label{fig:intro}
     \Description{}
\end{figure*}

Fig. \ref{fig:intro} illustrates this challenge with a simple four-qubit circuit.
The circuit is first translated into the ZX-diagram, where the original gate sequence is represented as a diagram of interacting Z and X spiders.
Prior work \cite{zhou2026topols} applies a restricted ZX reduction based on spider fusion while maintaining the structure that can be directly mapped to its lattice surgery junction model.
In this model, each spider is realized as a local lattice surgery junction, with its incident edges corresponding to connections to neighboring junctions; due to the planar geometry of a logical patch, a single junction supports at most four such connections.
Consequently, spider fusion is restricted to preserve the four-edge bound, rather than combined into the higher-degree spider exposed by full ZX reduction.
The resulting ZX diagram is then processed layer by layer, with Monte Carlo tree search \cite{coulom2006efficient} for the embedding of the junctions within each layer.
For this example, the resulting spacetime realization requires five time steps.

We present \emph{SpiderLS}, a lattice surgery compiler that further exploits the rewriting capabilities of the ZX intermediate representation.
Our key observation is that a reduced ZX-diagram need not preserve a one-to-one correspondence between each spider and a bounded-degree lattice surgery junction, since higher-arity interactions can instead be realized through multi-patch measurements \cite{litinski2019game}.
The lower half of Fig. \ref{fig:intro} illustrates this idea on the same example.
SpiderLS performs full ZX reduction \cite{van2020zx}, yielding a reduced diagram composed of Z spiders and blue ($H$) edges \cite{kissinger2019pyzx}, potentially with higher-degree spiders.

In this example, the reduced diagram contains a degree-five Z spider.
An execution order is then derived from the reduced diagram, where an $H$ edge connecting two Z spiders can be interpreted as a CZ operation \cite{van2020zx}.
Accordingly, edges 1--3 correspond to $\mathsf{CZ}(q_2,q_0);\mathsf{CZ}(q_2,q_3);\mathsf{CZ}(q_2,q_1)$, respectively.
Since these operations share the same logical qubit \(q_2\), they can be combined into a single-control multi-target operation, $\mathsf{CZ}^{(3)}(q_2;q_0,q_1,q_3)$.
The remaining edge 4 corresponds to a separate $\mathsf{CZ}(q_0,q_1)$, yielding a target program consisting of two operations. 
Thus, the original four operations are reduced to two.
SpiderLS subsequently lowers these operations to Pauli-product measurements, making both the participating patches and the required Pauli boundaries explicit. 
In this example, each operation is decomposed into two measurements. 
The two first-stage measurements can be scheduled concurrently, followed by the two dependent measurements. 
SpiderLS then routes each scheduled layer by selecting compatible Pauli boundaries for the participating patches and constructing routes that connect the required terminals, ultimately yielding a two-step spacetime realization.

We implement this approach and evaluate it against state-of-the-art circuit-based lattice surgery compilers (Liblsqecc \cite{watkins2024high} and DASCOT \cite{molavi2025dependency}) and a ZX-based compiler (TopoLS \cite{zhou2026topols}) using a diverse suite of algorithmic and randomly generated quantum circuits. 
SpiderLS consistently produces compact lattice surgery realizations while maintaining low compilation overhead. 
Compared with TopoLS, SpiderLS reduces spacetime volume by 49.2\% and compilation time by 99.8\%. 
Moreover, SpiderLS demonstrates strong scalability and achieves comparable or lower T-port density on Clifford+T workloads.

\section{Background}

We illustrate the surface code, lattice surgery, and ZX-calculus in this section, which are relevant to this work. For deeper knowledge of these topics, we recommend reading \cite{fowler2018low, litinski2019game} for surface code and lattice surgery and \cite{van2020zx} for ZX-calculus.

\subsection{Surface Code and Lattice Surgery}

The surface code is a quantum error correction code that encodes one logical qubit across a two-dimensional arrangement of multiple physical qubits and detects errors through repeated local measurements. 
It is implemented with data qubits, which store the encoded quantum information, and syndrome qubits, which interact with nearby data qubits to detect physical errors \cite{sethi2025rescq, zhu2026o3ls, ghosh2026toward, litinski2019game}.

\subsubsection{Physical Code}
Let $\mathcal{D}$ and $\mathcal{S}$ denote the sets of data and syndrome qubits, respectively.  
For each face $f$, let $\mathcal{N}(f) \subseteq \mathcal{D}$ denote its neighboring data qubits.  
The stabilizer associated with $f$ is
$$
\mathsf{stab}(f) =
\begin{cases}
  \displaystyle\prod_{q \in \mathcal{N}(f)} X_q,
    & f \text{ is an $X$ type (blue) face}, \\[6pt]
  \displaystyle\prod_{q \in \mathcal{N}(f)} Z_q,
    & f \text{ is a $Z$ type (red) face}.
\end{cases}
$$
Thus, $X$-type stabilizers detect $Z$ errors, and $Z$-type stabilizers detect $X$ errors. 
Fig. \ref{fig:surface_code_diagram} shows a $3\times3$ rotated surface code consisting of 9 data qubits and 8 syndrome qubits, and Fig. \ref{fig:stabilizers} shows the corresponding stabilizer measurement circuits.

\subsubsection{Logical Patches}\label{sec:logical_patches}
\begin{figure}[t]
     \centering
     \begin{subfigure}{0.17\textwidth}
         \centering
         \includegraphics[width=\textwidth]{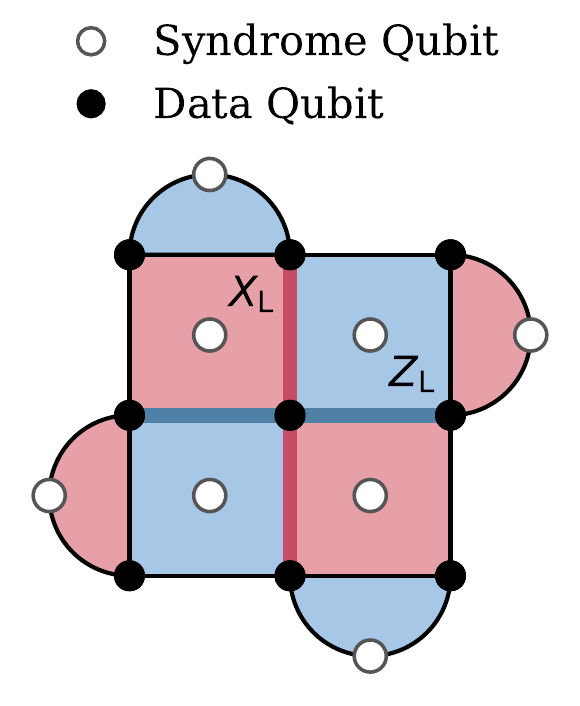}
         \caption{A 3 $\times$ 3 rotated surface code.}
         \label{fig:surface_code_diagram}
     \end{subfigure}
     \hfill
     \begin{subfigure}{0.29\textwidth}
         \centering
         \includegraphics[width=0.22\textwidth]{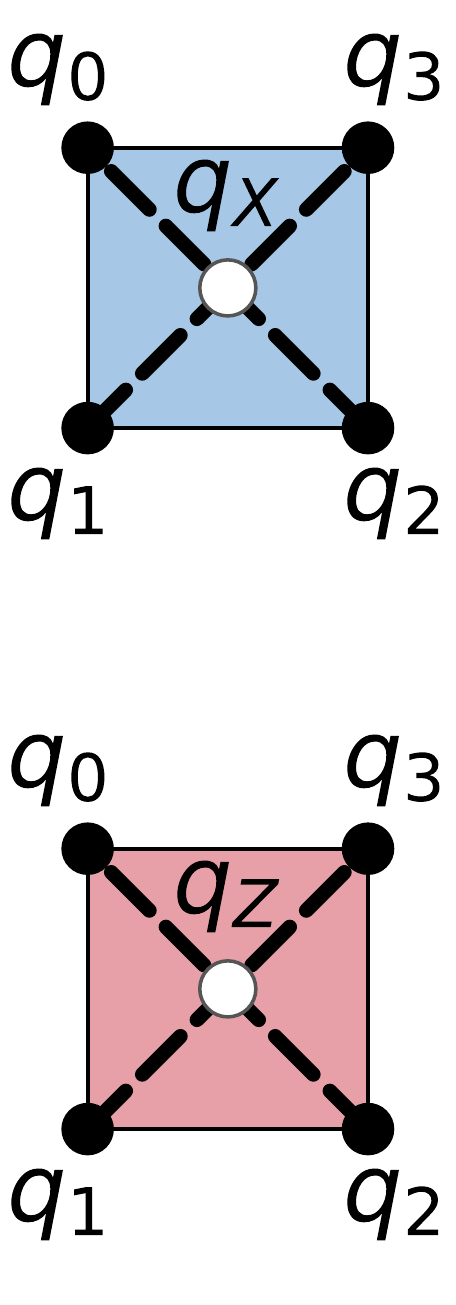}
         \includegraphics[width=0.76\textwidth]{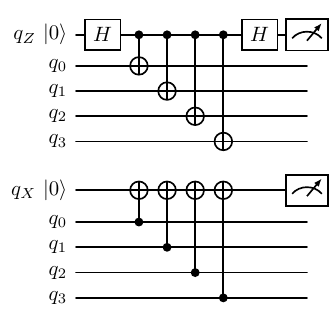}
         \caption{
         X-type (top) and Z-type (bottom) stabilizers, which detect Z and X errors, respectively.
         }
         \label{fig:stabilizers}
     \end{subfigure}
     \\
     \begin{subfigure}{0.42\textwidth}
         \centering
         \includegraphics[width=1\textwidth]{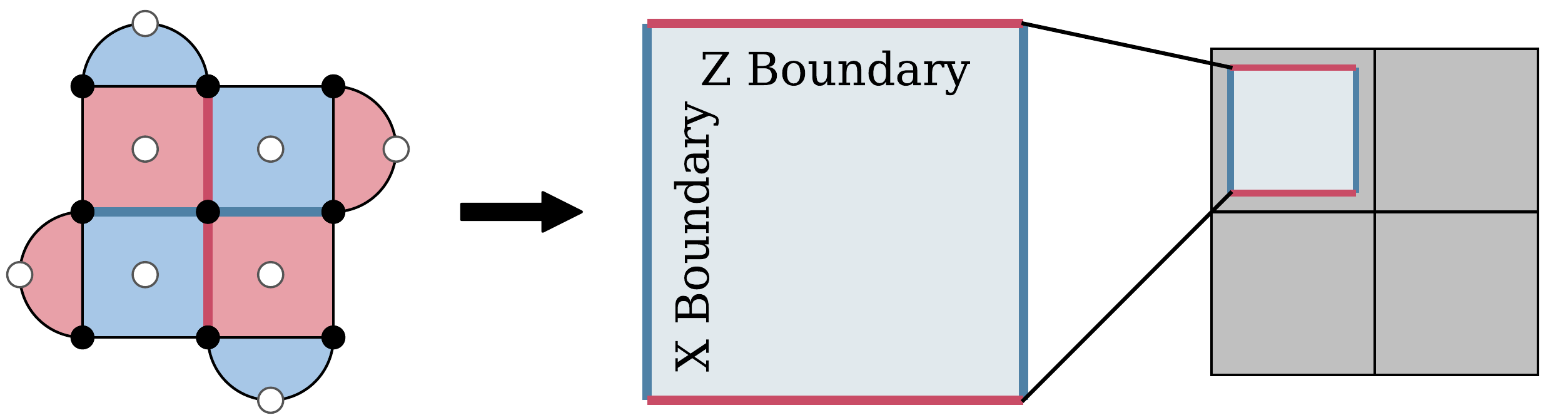}
         \caption{
         A surface code can be abstracted to a logical qubit patch that lies on the logical qubit tiles arranged in a grid.
         }
         \label{fig:surface_code_patch}
     \end{subfigure}
     \caption{Rotated surface code.}
     \label{fig:rotated_surface_code}
     \Description{}
\end{figure}

A surface code instance is abstracted as a \emph{logical patch}.
A patch has four directed boundaries,
$$
d ::= N \mid E \mid S \mid W,
$$
each of which has a boundary type
$$
b ::= X \mid Z.
$$
For a patch $P$, its boundary type map is written as
$$
\beta_P : \{N,E,S,W\} \rightarrow \{X,Z\}.
$$
We adopt the following orientation for a rotated patch representing logical qubit $q$:
$$
P_q
  \triangleq
  q[
    W{:}X,\;
    E{:}X,\;
    N{:}Z,\;
    S{:}Z
  ],
$$
or equivalently,
$$
\beta_{P_q}(W)
=
\beta_{P_q}(E)
=
X,
\qquad
\beta_{P_q}(N)
=
\beta_{P_q}(S)
=
Z.
$$
Fig. \ref{fig:surface_code_patch} illustrates this abstraction: the physical data and syndrome qubits are hidden, and computation is expressed in terms of logical patches arranged on a two-dimensional grid.

Logical Pauli operators are represented by chains of physical Pauli
operators connecting opposite boundaries.  
For paths
$\mathcal{P}_Z : W \xrightarrow{\text{path}} E$ and $\mathcal{P}_X : N \xrightarrow{\text{path}} S$,
$$
Z_L(P_q)
  =
  \prod_{v \in \mathcal{P}_Z} Z_v, 
\qquad
X_L(P_q)
  =
  \prod_{v \in \mathcal{P}_X} X_v.
$$
The corresponding logical $Z$ and $X$ basis measurements, $M_Z(P_q)$ and $M_X(P_q)$, are obtained by measuring the data qubits in the respective basis and decoding the logical outcome along these strings \cite{fowler2018low}.
Logical $Y$ basis measurement is implemented by a dedicated in-place fault-tolerant protocol within a surface code patch \cite{gidney2024inplace}, which we treat as an atomic single-patch operation $M_Y(P_q)$ \cite{tan2024sat}.

A domain wall operation \cite{litinski2018lattice} switches the $X$ and $Z$ boundary orientations of a patch,
$$
\beta_{P'}=\mathsf{swap}_{X,Z}(\beta_P),
$$
and thereby realizes a logical Hadamard, which exchanges $X_L$ and $Z_L$.

\subsubsection{Lattice Surgery}
\begin{figure}[t]
     \centering
     \begin{subfigure}{0.19\textwidth}
         \centering
         \includegraphics[width=\textwidth]{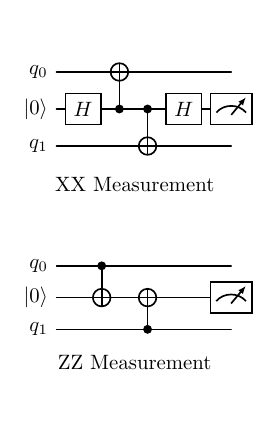}
         \caption{
         XX (top) and ZZ (bottom) measurements of $q_0$ and $q_1$ represented as quantum circuits.
         }
         \label{fig:measurements}
     \end{subfigure}
     \hfill
     \begin{subfigure}{0.25\textwidth}
         \centering
         \includegraphics[width=\textwidth]{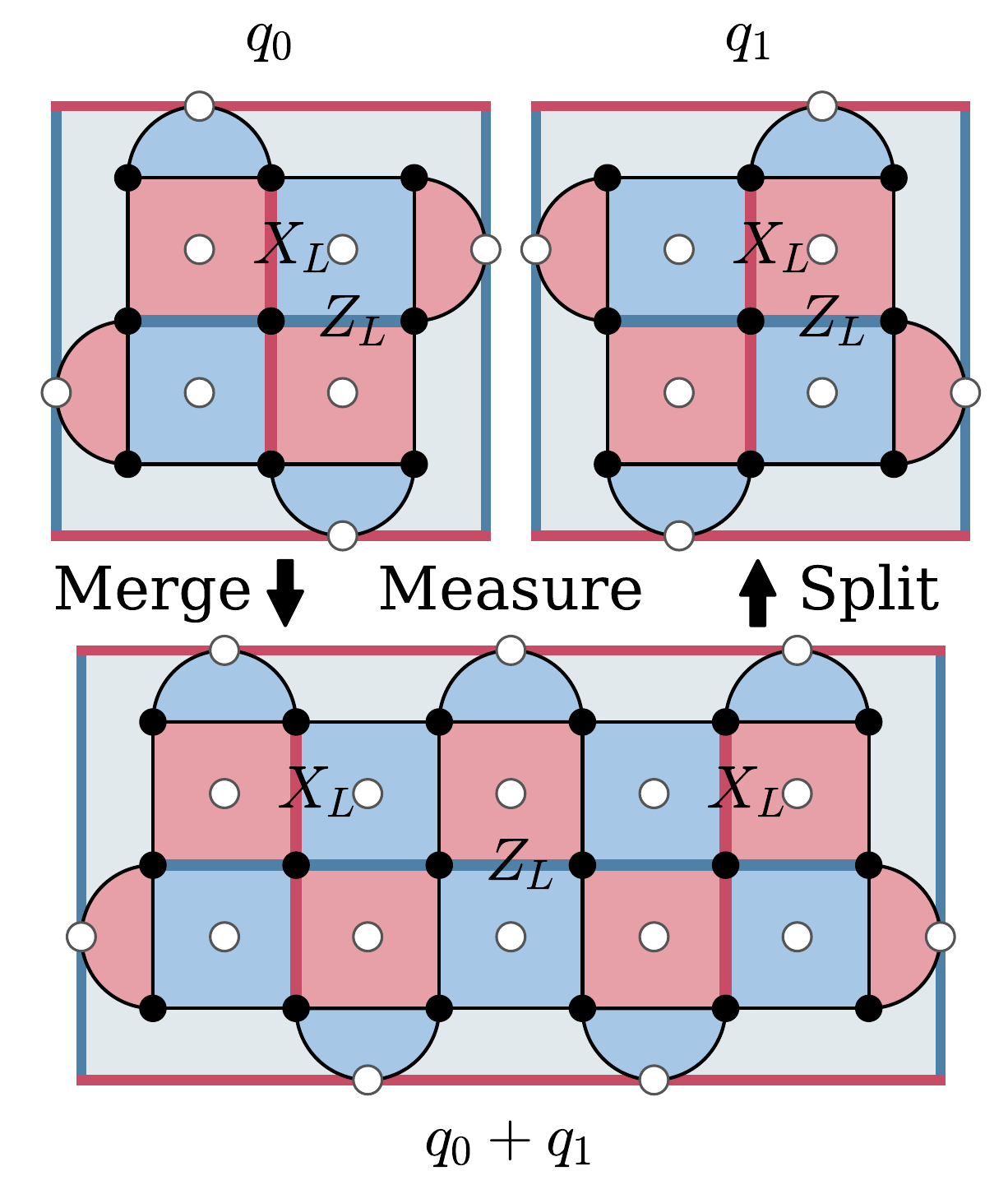}
         \caption{
         XX measurement implemented for lattice surgery. 
         The measurement result can be retrieved by merging the two patches.
         }
         \label{fig:surface_code_measurement}
     \end{subfigure}
     \caption{
     Pauli product measurements in quantum circuits and surface codes.
     }
     \label{fig:Pauli_product_measurements}
     \Description{}
\end{figure}

A Pauli-product measurement (PPM) \cite{litinski2019game} $M_{\mathbf p}(\mathbf q)$ jointly measures a Pauli product
$$
\mathbf p = p_1\cdots p_k,
\qquad
\mathbf q = (q_1,\ldots,q_k),
\qquad
p_i \in \{X,Z\},
$$
over $k\geq1$ logical qubits and produces a measurement outcome in $\{-1,+1\}$.
The $XX$ and $ZZ$ measurements shown in Fig. \ref{fig:measurements} are binary instances of PPMs.

At the surface code level, PPMs are realized through \emph{lattice surgery} \cite{horsman2012surface}, which performs PPMs by temporarily merging logical patches along appropriate boundaries and subsequently splitting them.
The boundary types participating in the merge determine the Pauli product being measured.
Fig. \ref{fig:surface_code_measurement} illustrates the lattice-surgery realization of an $XX$ measurement, where two logical patches are merged along their $X$ boundaries to obtain the measurement outcome.

\subsubsection{Gate Decomposition}
Logical gates can be decomposed into PPMs and single-qubit
measurements.  
We write
$$
G \Downarrow_{\mathrm{PPM}} \overline{M}
$$
when a logical gate $G$ is decomposed into a measurement sequence $\overline{M}$.  
As shown in Fig. \ref{fig:CXCZ}, CNOT and CZ each use an ancilla qubit $q_A$ initialized to $|0\rangle$, with
$$
\begin{aligned}
\mathsf{CNOT}(q_0,q_1)
\Downarrow_{\mathrm{PPM}}\;&
M_{XX}(q_1,q_A);
M_{ZZ}(q_0,q_A),
\\
\mathsf{CZ}(q_0,q_1)
\Downarrow_{\mathrm{PPM}}\;&
M_{ZX}(q_1,q_A);
M_{ZZ}(q_0,q_A),
\end{aligned}
$$
where the initialization and measurement of the ancilla qubit are omitted, as these operations do not require additional time steps \cite{litinski2019game}.
The measurement outcomes determine the Pauli corrections, which can be tracked classically using a Pauli frame rather than physically applied \cite{PennyLane-PauliFrameTracking, riesebos2017pauli}.

Fig. \ref{fig:lattice_surgery_cx} illustrates the lattice-surgery realization of CNOT for two time steps, where the $XX$ and $ZZ$ measurements are performed sequentially using three logical patches.

\subsubsection{Spacetime Diagram}
A lattice-surgery execution can also be represented as a three-dimensional spacetime diagram, where the two spatial dimensions---$x$ and $y$ axes---encode the patch layout and the third dimension---$z$ axis---represents time \cite{gidney2019flexible}.
Under this view, a logical patch traces out a volume through time, while merge and split operations appear as changes in the connectivity of these volumes.
It is defined that $\text{(spacetime volume)} = \text{(area)} \times \text{(time steps)}$, which indicates the required resources \cite{tan2024sat}.

\begin{figure}[t]
     \centering
     \begin{subfigure}{0.48\textwidth}
         \centering
         \includegraphics[width=\textwidth]{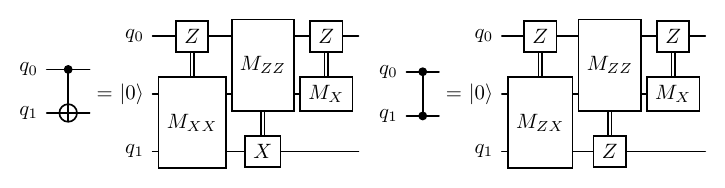}
         \caption{
         Lattice surgery implementations of CNOT (left) and CZ (right). Each implementation requires an ancilla patch.
         }
         \label{fig:CXCZ}
     \end{subfigure}
     \\
     \begin{subfigure}{0.27\textwidth}
         \centering
         \includegraphics[width=0.97\textwidth]{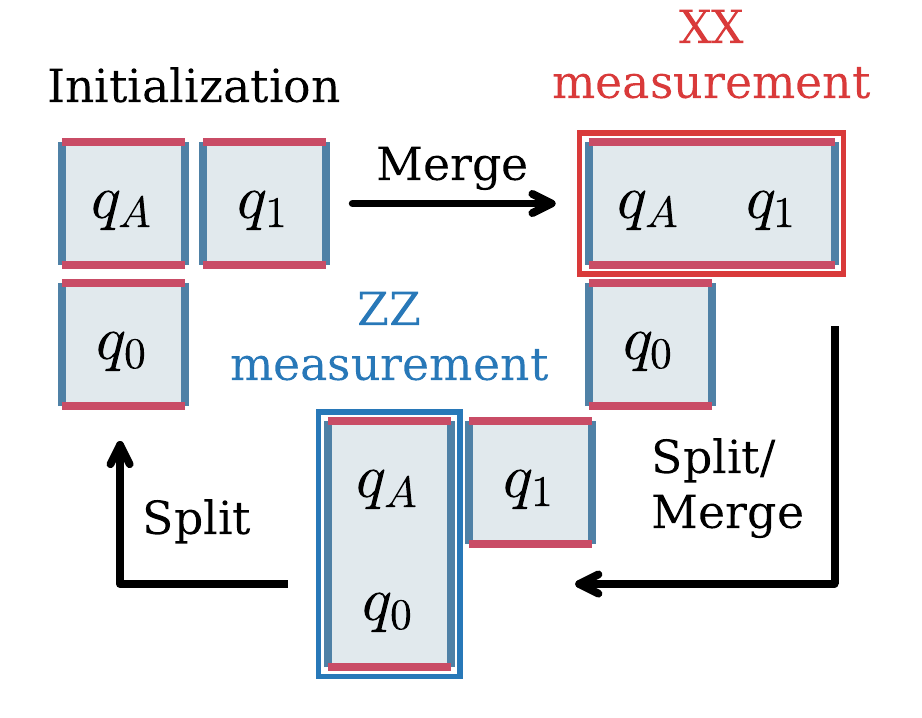}
         \caption{
         CNOT gate implemented as lattice surgery operations. Three patches and two time steps are required.
         }
         \label{fig:lattice_surgery_cx}
     \end{subfigure}
     \hfill
     \begin{subfigure}{0.20\textwidth}
         \centering
         \includegraphics[width=\textwidth]{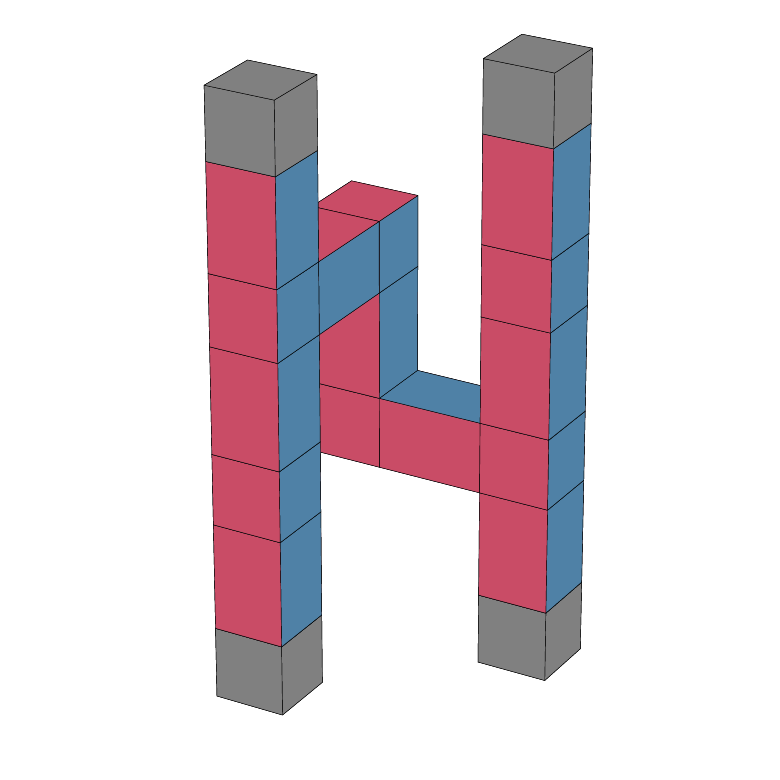}
         \caption{
         Lattice surgery operations of CNOT represented as a spacetime diagram.
         }
         \label{fig:pipe_cnot}
     \end{subfigure}
     \caption{
     Controlled gate implemented by lattice surgery.
     }
     \label{fig:controlled_gates}
     \Description{}
\end{figure}

Fig. \ref{fig:pipe_cnot} represents the same execution as Fig. \ref{fig:lattice_surgery_cx}.  
The gray ports indicate inputs and outputs, and each floor represents a time step.
The spacetime volume is $x \cdot y \cdot z = 2 \cdot 2 \cdot 2 = 8$.

\subsection{ZX-calculus}

\begin{figure}[b]
     \centering
     \begin{subfigure}{0.38\textwidth}
         \centering
         \includegraphics[width=\textwidth]{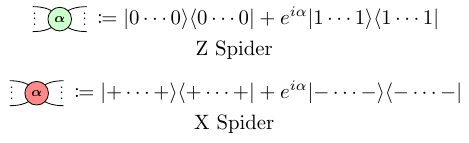}
         \caption{
         Definition of Z and X spiders with phase $\alpha$. Phase 0 can be omitted.
         }
         \label{fig:spiders}
     \end{subfigure}
     \\
     \begin{subfigure}{0.48\textwidth}
         \centering
         \includegraphics[width=0.27\textwidth]{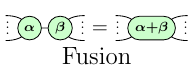}
         \includegraphics[width=0.24\textwidth]{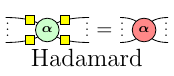}
         \includegraphics[width=0.12\textwidth]{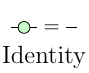}
         \includegraphics[width=0.15\textwidth]{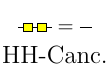}
         \includegraphics[width=0.25\textwidth]{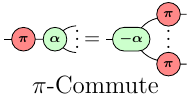}
         \includegraphics[width=0.16\textwidth]{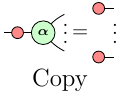}
         \includegraphics[width=0.17\textwidth]{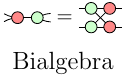}
         \includegraphics[width=0.19\textwidth]{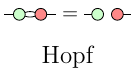}
         \caption{
         The major graphical rewrite rules of ZX-calculus.
         }
         \label{fig:zx_rule}
     \end{subfigure}
     \caption{
     Basic concepts of ZX-calculus.
     }
     \label{fig:zx_spiders}
     \Description{}
\end{figure}

ZX-calculus \cite{coecke2011interacting} is a graphical language for linear maps and can be used as an \emph{intermediate representation (IR)} of quantum programs \cite{zhou2026topols, PennyLane-ZX-Calculus}.
A ZX-diagram is composed of Z and X spiders, Hadamard nodes, and sequential and parallel compositions:
$$
D ::= Z_{\alpha}^{m,n}
\mid X_{\alpha}^{m,n}
\mid H
\mid D \circ D
\mid D \otimes D .
$$
where $m$ and $n$ are the numbers of inputs and outputs, respectively.
Each diagram $D$ denotes a linear map $\llbracket D \rrbracket$.
Fig. \ref{fig:spiders} gives the semantics of Z and X spiders; in particular, spiders without inputs or outputs represent initializations or measurements, respectively;
\begin{equation*}
    \raisebox{-0.9ex}{\includegraphics[height=1.4em]{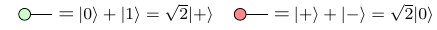}},
\end{equation*}
\begin{equation*}
    \raisebox{-0.9ex}{\includegraphics[height=1.4em]{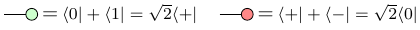}}.
\end{equation*}

ZX-diagrams admit local, semantics-preserving rewrite rules.
We write
$$
D \longrightarrow_{\mathrm{ZX}}^{*} D'
\quad\Longrightarrow\quad
\llbracket D \rrbracket = \llbracket D' \rrbracket
$$
for a sequence of such rewrites, up to an ignored global scalar.
The major rules are summarized in Fig. \ref{fig:zx_rule}, where the detailed proofs can be found in \cite{van2020zx}.

Utilizing these properties, quantum circuits can be translated to ZX-diagrams.
We denote $\mathcal{T}_{\mathrm{ZX}}(C)$ as the translation of a quantum circuit $C$ into a ZX-diagram.
For example, CNOT is translated into its corresponding ZX-diagram as follows:
{\small
\begin{align*}
    \raisebox{-2.4ex}{\includegraphics[height=2.5em]{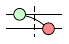}}
    &= \Bigl(I \otimes (|+\rangle\langle++|+|-\rangle\langle--|)\Bigr)\Bigl((|00\rangle\langle0|+|11\rangle\langle1|) \otimes I\Bigr)\\
    &= \frac{1}{\sqrt{2}}
    \left( \begin{smallmatrix}
    1 & 0 & 0 & 0\\
    0 & 1 & 0 & 0\\
    0 & 0 & 0 & 1\\
    0 & 0 & 1 & 0
    \end{smallmatrix} \right) \cong
    \raisebox{-2.4ex}{\includegraphics[height=2.5em]{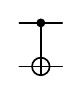}}
\end{align*}
}
Using the translation, the PPM circuits in Fig. \ref{fig:measurements}, $M_{XX}$ and $M_{ZZ}$, can be translated as
\begin{equation*}
    \raisebox{-3ex}{\includegraphics[height=3em]{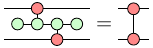}} \quad \text{and} \quad \raisebox{-3ex}{\includegraphics[height=3em]{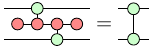}},
\end{equation*}
respectively.
Therefore, we can now draw the ZX-diagram of the lattice surgery implementation of CNOT in Fig. \ref{fig:CXCZ}:
\begin{equation*}
    \raisebox{-3ex}{\includegraphics[height=3em]{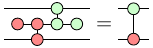}},
\end{equation*}
which matches the previous diagram.

Additionally, since $\text{CZ} = (I \otimes H) \cdot \text{CNOT} \cdot (I \otimes H)$, CZ can be translated as follows:
\begin{equation*}
    \raisebox{-3ex}{\includegraphics[height=3em]{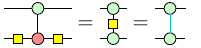}},
\end{equation*}
where an $H$ gate can also be represented as a blue ($H$) edge.

\section{Motivation}
\label{sec:motivation}

This section introduces the state-of-the-art work \cite{zhou2026topols} that utilizes ZX-calculus as an IR in lattice surgery compilation and opportunities for further exploiting ZX reduction.

\subsection{ZX-Based Lattice Surgery Compilation}

The state-of-the-art work \cite{zhou2026topols} uses ZX-calculus as an IR for lattice-surgery compilation.  
Its compilation pipeline can be summarized as
$$
C
\xrightarrow{\mathcal{T}_{\mathrm{ZX}}}
D
\xrightarrow{\mathsf{fuse}}
D_f
\xrightarrow{\mathsf{slice}}
(D_f,\lambda)
\xrightarrow{\mathsf{embed}}
S ,
$$
where $C$ is the input circuit, $D_f$ is a simplified ZX diagram using the Fusion rule in Fig. \ref{fig:zx_rule}, $\lambda: V(D_f)\rightarrow\mathbb{N}$ assigns spiders to execution layers, and $S$ is a lattice surgery spacetime diagram.
Spider fusion performs semantics-preserving IR simplification, whereas the final embedding pass determines the geometric placement and routing of the resulting graph in spacetime using Monte Carlo tree search (MCTS) \cite{coulom2006efficient}.

This approach relies on the correspondence between ZX spiders and lattice surgery junctions.
A spider represents a local interaction point in the IR, and its incident edges must be realized by a single logical patch in the spacetime diagram.
Suppose that several PPM interactions are incident on the same patch $P$, and let $p_i(P)\in\{X, Z\}$ denote the Pauli operator required from $P$ by the $i$th interaction.
To realize them simultaneously at one junction, they must be locally compatible \cite{litinski2019game}:
$$
[p_i(P),p_j(P)] = 0 .
$$
Since
$$
[X,X]=[Z,Z]=0,
\qquad
[X, Z]\neq 0,
$$
all interactions incident on a single square patch junction must therefore use the same Pauli type.
Hence, once the Pauli type is fixed, the patch can participate only through one pair of opposite spatial boundaries:
$$
B_X=\{E,W\},
\qquad
B_Z=\{N, S\}.
$$
A spacetime junction additionally has one incoming and one outgoing temporal direction, denoted by $t^{-}$ and $t^{+}$.
Therefore, the set of available ports of a single junction is bounded by
$$
\operatorname{ports}(v)
\subseteq
B_p\cup\{t^{-},t^{+}\},
\qquad
|\operatorname{ports}(v)|\le 4 .
$$
Since each incident port is represented as an edge of the corresponding ZX spider, the following IR-level invariant can be derived:
\begin{equation}\label{lb:deg}
\deg(v)\le 4
\qquad
\text{for every spider } v\in V(D_f).
\end{equation}

\begin{figure}[ht]
     \centering
     \begin{subfigure}{0.15\textwidth}
         \centering
         \includegraphics[width=\textwidth]{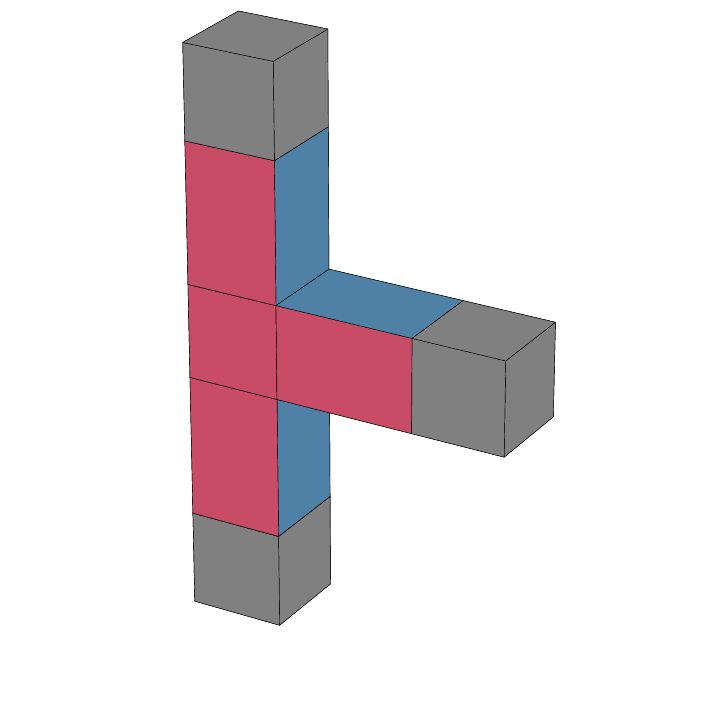}
         \caption{One connected patch; a three-edge spider (valid).}
         \label{fig:pipe3}
     \end{subfigure}
     \hfill
     \begin{subfigure}{0.15\textwidth}
         \centering
         \includegraphics[width=\textwidth]{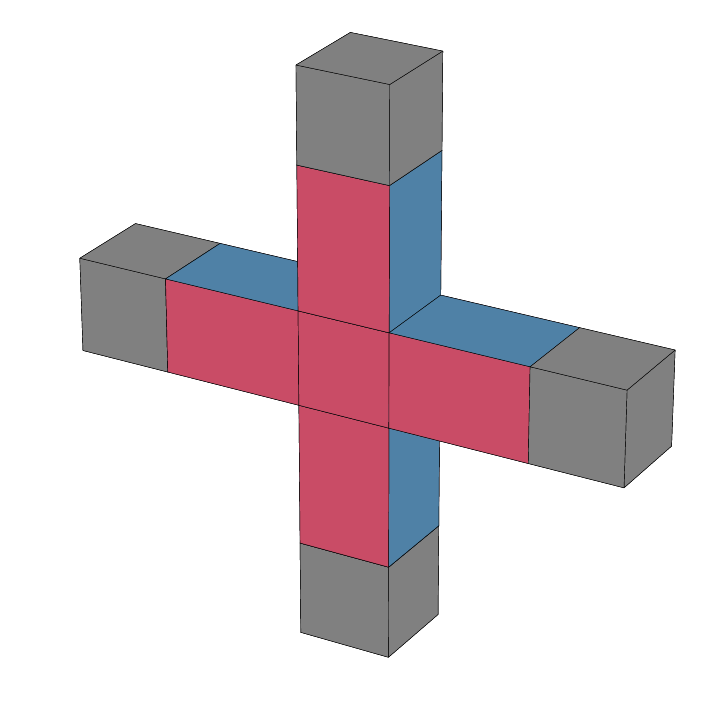}
         \caption{Two connected patches; a four-edge spider (valid).}
         \label{fig:pipe4}
     \end{subfigure}
     \hfill
     \begin{subfigure}{0.15\textwidth}
         \centering
         \includegraphics[width=\textwidth]{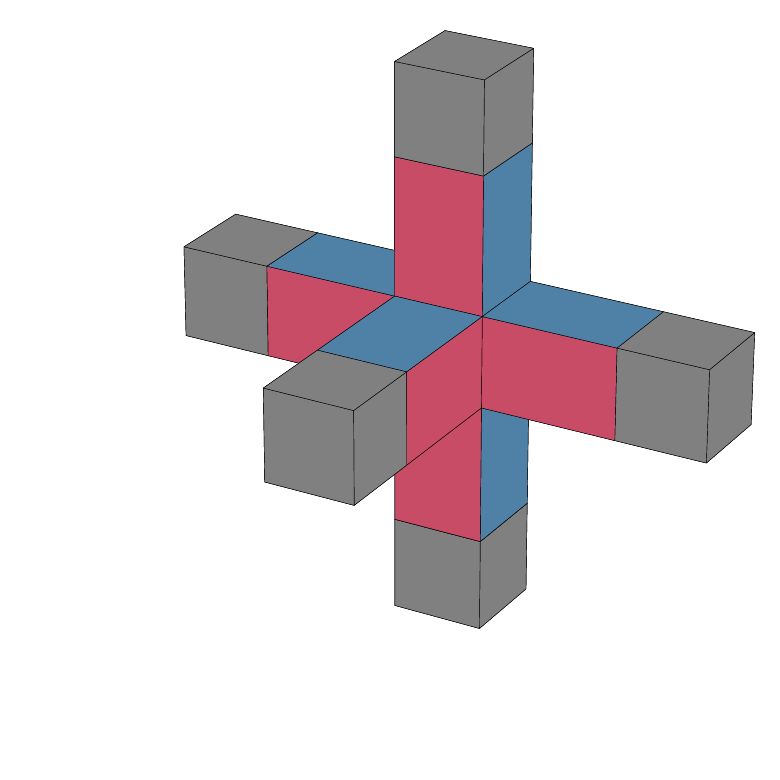}
         \caption{Three connected patches; a five-edge spider (invalid).}
         \label{fig:pipe5}
     \end{subfigure}
     \caption{Validity of a lattice surgery junction as a function of the number of patches connected to a single logical patch.}
     \label{fig:pipe_n}
     \Description{}
\end{figure}

Fig. \ref{fig:pipe_n} illustrates this constraint in a graphical manner.
If one additional patch is connected to a logical patch, the corresponding spider has three edges: two temporal edges and one spatial edge (Fig. \ref{fig:pipe3}).
If two patches are connected, the spider has four edges and remains valid (Fig. \ref{fig:pipe4}).
However, connecting three patches requires five edges---two temporal and three spatial---which violates the constraint and is therefore unrealizable as a single junction (Fig. \ref{fig:pipe5}).

\subsection{Multi-Patch Measurements for Single-Control Multiple-Target Gates}
\label{sec:motivation2}

The preceding approach, however, leaves an opportunity to distinguish an important distinction between representing a ZX spider as a single pipe junction and implementing the corresponding logical interaction by lattice surgery.
While the former is subject to the four-port constraint, lattice surgery can realize higher-arity Pauli-product measurements through multi-patch measurements.

Consider a single-control, $k$-target CZ operation:
$$
\mathsf{CZ}^{(k)}
(q_c;q_1,\ldots,q_k)
\triangleq
\prod_{i=1}^{k}\mathsf{CZ}(q_c,q_i).
$$
The constituent CZ gates share the control qubit $q_c$.
After translating the circuit into ZX-calculus and fusing the spiders along the shared control, we obtain
$$
\mathcal{T}_{\mathrm{ZX}}
\!\left(\mathsf{CZ}^{(k)}\right)
\longrightarrow_{\mathrm{ZX}}^{*}
D_k,
\qquad
\deg_{D_k}(v_c)=k+2,
$$
where $v_c$ denotes the fused control spider and the additional two edges correspond to the input and output of $q_c$.
Under the four-edge constraint in Eq. \ref{lb:deg}, the fused spider is therefore directly realizable as a single junction only if
$$
k+2\le4
\quad\Longrightarrow\quad
k\le2.
$$

We illustrate this degree growth for $k=1,2,$ and $3$.
For $k=1$, $\mathrm{CZ}^{(1)}$ is an ordinary two-qubit CZ gate.
Its ZX translation contains a control spider of degree three and therefore satisfies Eq. \ref{lb:deg}:
\begin{align*}
    \raisebox{-2.5ex}{\includegraphics[height=2.5em]{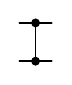}}
    \xrightarrow{\mathcal{T}_{\mathrm{ZX}}}
    \raisebox{-2.5ex}{\includegraphics[height=2.5em]{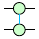}}.
\end{align*}
For $k=2$, the two CZ gates share the same control qubit.
After translation, spider fusion combines the interactions on the shared control into a single spider of degree four:
\begin{align*}
    \raisebox{-4.0ex}{\includegraphics[height=4.0em]{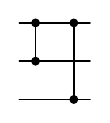}}
    \xrightarrow{\mathcal{T}_{\mathrm{ZX}}}
    \raisebox{-4.0ex}{\includegraphics[height=4.0em]{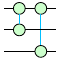}}
    \longrightarrow_{\mathrm{ZX}}^{*}
    \raisebox{-4.0ex}{\includegraphics[height=4.0em]{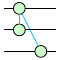}}.
\end{align*}
The resulting diagram is still admissible under Eq. \ref{lb:deg}.
For $k=3$, the same semantics-preserving fusion produces a degree-five control spider:
\begin{align*}
    \raisebox{-5.8ex}{\includegraphics[height=5.5em]{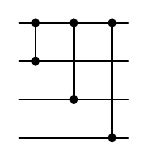}}
    \xrightarrow{\mathcal{T}_{\mathrm{ZX}}}
    \raisebox{-5.8ex}{\includegraphics[height=5.5em]{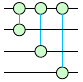}}
    \longrightarrow_{\mathrm{ZX}}^{*}
    \raisebox{-5.8ex}{\includegraphics[height=5.5em]{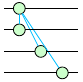}}.
\end{align*}
Hence, the fused diagram violates Eq. \ref{lb:deg} and cannot be directly instantiated as a single four-port junction.

Crucially, this violation does not imply that the three-target CZ is unrealizable by lattice surgery.
The degree bound constrains the representation of a spider as a single junction, whereas lattice surgery also supports \emph{multi-patch measurements}, in which a Pauli product is measured jointly across more than two logical patches \cite{zhu2026o3ls, litinski2019game}.

Indeed, all three examples above can be decomposed into PPMs and single-qubit measurements.
Let $q_A$ be an ancillary logical qubit initialized to $|0\rangle$.
Then
$$
\mathsf{CZ}^{(k)}\!(q_c;q_1,\!\ldots\!,q_k)\!
\Downarrow_{\mathrm{PPM}}
M_{Z^kX}\!(q_1,\!\ldots\!,q_k,q_A);
M_{ZZ}\!(q_c,q_A)
$$
up to the outcome-dependent Pauli-frame corrections \cite{fowler2018low}.
For example,
$$
\begin{aligned}
\mathsf{CZ}^{(1)}
&\Downarrow_{\mathrm{PPM}}
M_{ZX}(q_1,q_A);
M_{ZZ}(q_c,q_A),\\
\mathsf{CZ}^{(2)}
&\Downarrow_{\mathrm{PPM}}
M_{ZZX}(q_1,q_2,q_A);
M_{ZZ}(q_c,q_A),\\
\mathsf{CZ}^{(3)}
&\Downarrow_{\mathrm{PPM}}
M_{ZZZX}(q_1,q_2,q_3,q_A);
M_{ZZ}(q_c,q_A).
\end{aligned}
$$
The important distinction is that increasing the number of targets increases the \emph{arity} of the first PPM rather than the number of ancillary logical qubits or measurement stages.
In particular, the three-target CZ uses the four-patch measurement $M_{ZZZX}(q_1,q_2,q_3,q_A)$ and is therefore realizable even though its fused ZX representation contains a degree-five spider.

\begin{figure}[ht]
     \centering
     \begin{subfigure}{0.15\textwidth}
         \centering
         \includegraphics[width=\textwidth]{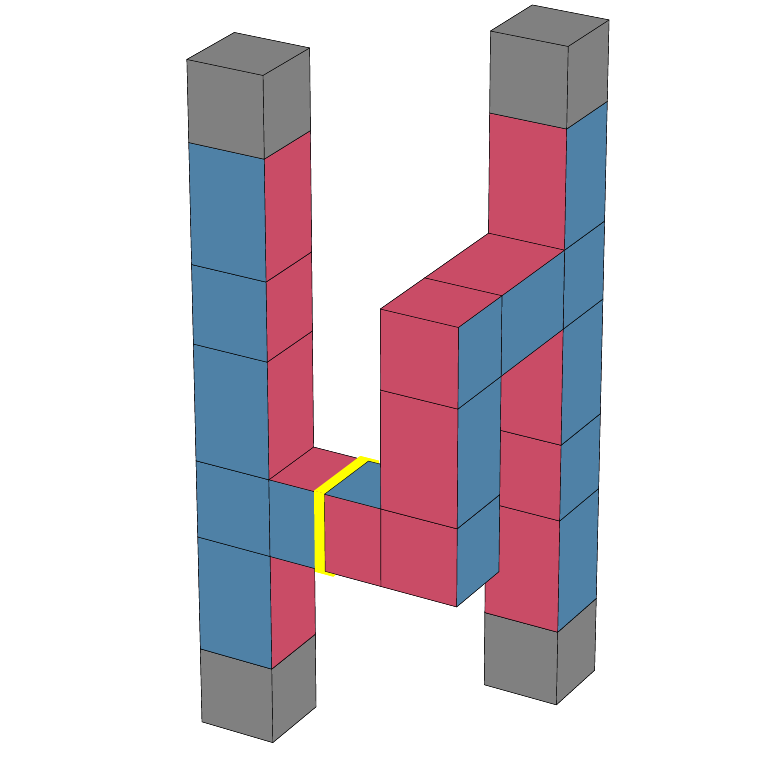}
         \caption{Single-target CZ.}
         \label{fig:cz1}
     \end{subfigure}
     \hfill
     \begin{subfigure}{0.15\textwidth}
         \centering
         \includegraphics[width=\textwidth]{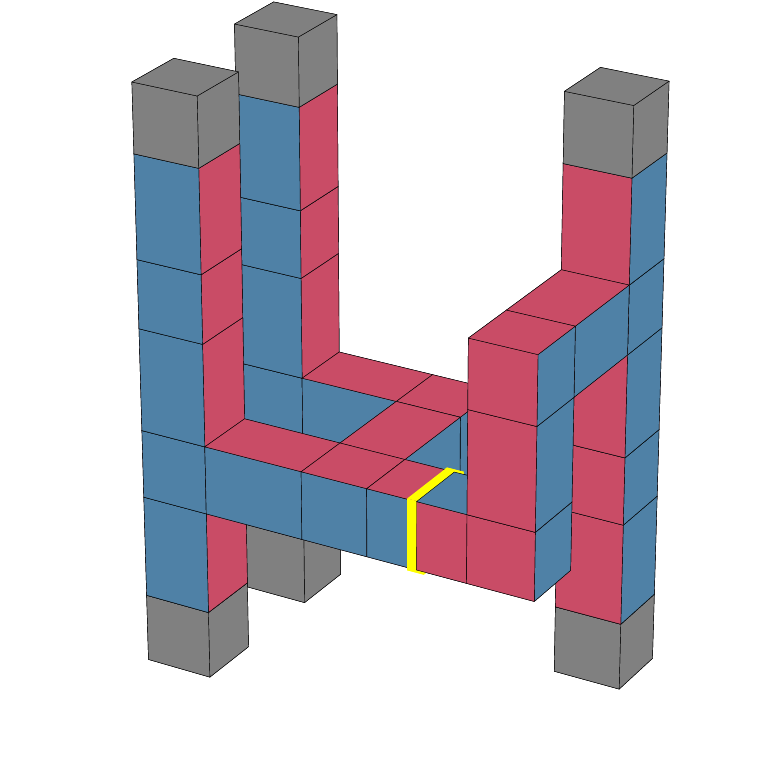}
         \caption{Two-target CZ.}
         \label{fig:cz2}
     \end{subfigure}
     \hfill
     \begin{subfigure}{0.15\textwidth}
         \centering
         \includegraphics[width=\textwidth]{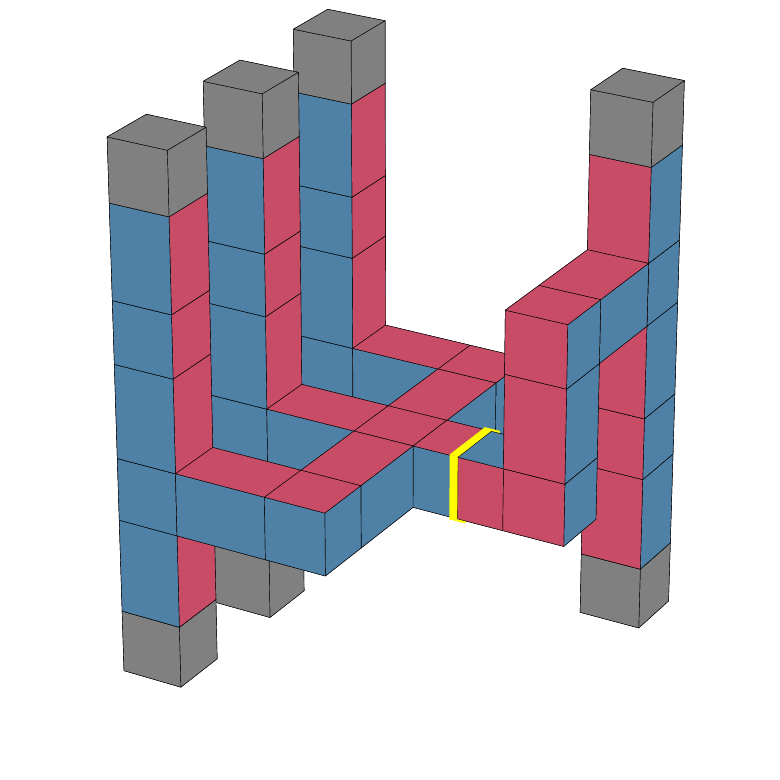}
         \caption{Three-target CZ.}
         \label{fig:cz3}
     \end{subfigure}
     \caption{Spacetime diagrams of single-control multiple-target CZ gates implemented by multi-patch measurements.}
     \label{fig:cz_n}
     \Description{}
\end{figure}

Fig. \ref{fig:cz_n} shows the corresponding spacetime diagrams for $k=1,2,$ and $3$.
The three-target case in Fig. \ref{fig:cz3} provides a concrete example of the gap between the degree constraint of the junction-based representation and the actual expressiveness of lattice surgery:
$$
\deg(v)>4
\;\not\Rightarrow\;
\text{unimplementable by lattice surgery}.
$$
Rather, it means that the interaction cannot be represented directly as a single four-port junction.
Thus, ZX reduction need not be restricted merely to satisfy the four-port junction constraint; they can instead be lowered directly to higher-arity multi-patch measurements.

\subsection{Structure-Aware Spacetime Routing}

Preceding work \cite{zhou2026topols} reduces the global embedding problem by first slicing the ZX-diagram into execution layers and then embedding the layers sequentially.  
Within each layer, however, placement, orientation, and routing decisions remain coupled: MCTS incrementally embeds the spiders in the current layer while respecting the geometry fixed by preceding layers. 
Once a layer is completed, its embedding is fixed and used as the basis for subsequent layers.
Thus, for a layer $L$ with spiders $V_L$, its geometric choices can be viewed as a joint design space:
$$
\mathcal A_L
\subseteq
\prod_{v\in V_L}\mathcal A_v ,
$$
which is explored by MCTS without exhaustively enumerating all combinations.

SpiderLS exploits stronger structure exposed after PPM scheduling.
Each measurement $m=M_{\mathbf p}(\mathbf q)$ already identifies the participating qubits $\mathbf q$ and the required Pauli boundary at each terminal.  
Once earlier routing decisions have been committed, routing therefore reduces to a local problem conditioned on the current geometry:
$$
\langle m,\Omega\rangle
\Downarrow_{\mathrm{route}}
\rho,
$$
where $\Omega$ contains the currently fixed patch placements and routes, and $\rho$ is a geometric realization of $m$.

Let $M_{Rt}=(m_1,\ldots,m_s)$ be the PPMs considered in routing layer $t$.  
After routing the first $j-1$ measurements, SpiderLS searches only a bounded local candidate set:
$$
\mathcal B(\!m_j,\!\Omega_{t,j}\!)
\!=\!
\{\!\rho \!\mid\!
\rho \!\text{ realizes }\! m_j
\text{ and is compatible with }\! \Omega_{t,j}\!\}.
$$
If each local search examines at most $\nu$ candidates, the candidate work is organized as
$$
\sum_{j=1}^{s}
\left|\mathcal B(m_j,\Omega_{t,j})\right|
\le \nu s,
$$
rather than as a search tree over combinations of routing decisions for the entire layer.  
This does not imply globally optimal routing; instead, SpiderLS deliberately replaces joint layer-level optimization with bounded conditional searches.

A successful route is committed immediately and becomes part of the geometry seen by subsequent PPMs.  
If no compatible route is found within the bounded search, the measurement is deferred to a later physical layer, while other independent ready measurements may still proceed.
This commit-or-defer strategy uses physical time as a spill resource for spatial conflicts, avoiding an MCTS search over the embedding choices of an entire layer.

\section{SpiderLS}

SpiderLS is a compilation framework designed to transform Clifford+T circuits into lattice surgery spacetime realizations via a sequence of intermediate representations:
$$
\begin{aligned}
& C
\xrightarrow{\mathcal T_{\mathrm{ZX}}}
D
\longrightarrow_{\mathrm{ZX}}^{*}
D_r
\xrightarrow{\mathsf{order}}
\pi\\
& \xrightarrow{\mathsf{group}}
\overline G_{\mathrm{tgt}}
\Downarrow_{\mathrm{PPM}}
\overline M
\xrightarrow{\mathsf{schedule}}
\Sigma
\xrightarrow{\mathsf{route}}
S.
\end{aligned}
$$
Here, $C$ is the input circuit, $D$ its ZX representation, $D_r$ the fully reduced ZX graph, and $\pi$ an execution order extracted from $D_r$.
The ordered interactions are grouped into a target program $\overline G_{\mathrm{tgt}}$, lowered to a PPM sequence $\overline M$, and scheduled into logical layers $\Sigma$.
Finally, SpiderLS maps $\Sigma$ to a lattice-surgery spacetime realization $S$.

The design of SpiderLS follows the two observations from Sec. \ref{sec:motivation}.
First, because high-degree ZX interactions can be realized through multi-patch measurements, SpiderLS allows unrestricted ZX reduction and utilizes single-control multi-target operations for reducing the length of the target code.
Second, once these operations are lowered to PPMs, the participating patches and required Pauli boundaries are explicit.
SpiderLS therefore constructs the spacetime realization through layer-wise local routing.

The remainder of this section describes the four main stages of SpiderLS: (1) ZX reduction and execution ordering, (2) target-code generation, (3) PPM lowering and logical scheduling, and (4) layered spacetime routing.

\begin{figure}[ht]
     \centering
     \begin{subfigure}{0.2\textwidth}
         \centering
         \includegraphics[width=\textwidth]{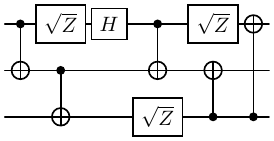}
         \caption{An example input circuit.}
         \label{fig:sample}
     \end{subfigure}
     \hfill
     \begin{subfigure}{0.265\textwidth}
         \centering
         \includegraphics[width=\textwidth]{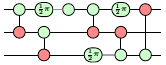}
         \caption{Translated to a ZX-diagram.}
         \label{fig:sample_zx}
     \end{subfigure}
     \\
     \begin{subfigure}{0.4\textwidth}
         \centering
         \includegraphics[width=\textwidth]{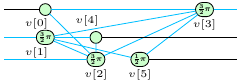}
         \caption{ZX-diagram after full reduction. Six Z spiders are connected with nine H edges.}
         \label{fig:sample_zx_reduce}
     \end{subfigure}
     \caption{ZX translation and full reduction of an example input circuit.}
     \label{fig:reduction_ordering}
     \Description{}
\end{figure}

\subsection{ZX Reduction and Execution Ordering}

SpiderLS first translates the input Clifford+T circuit $C$ into a ZX-diagram using PyZX \cite{kissinger2019pyzx}, 
$$
C \xrightarrow{\mathcal T_{\mathrm{ZX}}} D. 
$$
Fig. \ref{fig:sample} shows an example input circuit, and Fig. \ref{fig:sample_zx} shows its corresponding ZX representation.

SpiderLS then applies PyZX's \texttt{full\_reduce()} routine:
$$
D \longrightarrow_{\mathrm{ZX}}^{*} D_r, \qquad \llbracket D \rrbracket \simeq \llbracket D_r \rrbracket , 
$$
where $\simeq$ denotes equality up to an ignored global scalar. 
The original purpose of \texttt{full\_reduce()} is to expose and reduce non-Clifford structure, in particular the T count, which is achieved through the following steps \cite{kissinger2019pyzx, PennyLane-ZX-Calculus}:
\begin{enumerate}
    \item The input circuit is translated to a \emph{graph-like normal form}, which consists of Z spiders and H edges.
    \item Clifford structure is eliminated using the rewrite rules in Fig. \ref{fig:zx_rule}.
    \item The remaining Pauli structure is transformed into phase gadgets, and the compatible gadgets are fused.
    \item Steps 2 and 3 are repeated until no further reduction opportunity remains.
\end{enumerate}
Fig. \ref{fig:sample_zx_reduce} shows the fully reduced graph of the running example.

Full reduction no longer preserves the sequential structure of the input circuit. 
SpiderLS therefore derives an execution order directly from the reduced ZX graph. 
Let $D_r=(V, E,\phi)$, where $V$ is the set of internal Z spiders, $E$ is the set of H edges, and $\phi(v)$ is the phase of spider $v$. 
The ordering pass does not modify $\phi$; the phases are preserved as metadata for target code generation.

At step $j$, the ordering state is
$$
\sigma_j
=
\langle N_j,R_j,\mu_j,\pi_j\rangle ,
$$
where $N_j\subseteq V$ is the set of active spider locations, $R_j\subseteq E$ is the set of unprocessed edges, $\mu_j$ maps each live logical data patch $q$ to its current spider, and $\pi_j$ is the execution sequence emitted so far.
Initially,
$$
N_0=V_{\mathrm{in}},
\quad
R_0=E,
\quad
\mu_0(q)=v\in V_{\mathrm{in}},
\quad
\pi_0=[\,],
$$
where $V_{\mathrm{in}}$ contains the spiders adjacent to the input edges, and the initial number of live data patches is equal to the number of input edges.

SpiderLS updates this state using four rules.
First, interactions whose two endpoints are active are executable:
$$
\mathcal X_j
=
\left\{
\{u,v\}\in R_j
\mid
u,v\in N_j
\right\},
$$
and the edges are emitted as CZs and removed from $R_j$.
Second, an active non-output spider with no remaining incident edge has completed its lifetime:
$$
\mathcal F_j
=
\left\{
u\in N_j\setminus V_{\mathrm{out}}
\mid
\deg_{R_j}(u)=0
\right\},
$$
and is removed from the active set, measuring the data patch in X basis.
The measurement outcome is retained classically and its induced Pauli correction is tracked in the Pauli frame \cite{PennyLane-PauliFrameTracking, riesebos2017pauli}.
Third, an active non-output spider with exactly one remaining edge may advance to its inactive neighbor:
$$
\mathcal H_j\!
=\!
\left\{\!
(u,v)\!
\!\;\middle|\;\!
u\!\in \!N_j\!\setminus \!V_{\mathrm{out}},
\deg_{R_j}\!(u)\!=\!1,
\{u,v\}\!\in\! R_j,
v\!\notin\! N_j\!
\right\}\!.
$$
For $(u,v)\in\mathcal H_j$ and $\mu_j(q)=u$, $H$ is emitted and updated as
$$
\mu_{j+1}(q)=v,
\qquad
N_{j+1}=(N_j\setminus\{u\})\cup\{v\}.
$$
Finally, if unprocessed edges remain but neither an executable interaction nor an H move is available, the active spider set is expanded.
For each inactive spider $v$, define
$$
\eta_j(v)
=
\left|
\left\{
\{u,v\}\in R_j
\mid
u\in N_j
\right\}
\right|.
$$
SpiderLS activates
$$
v^*
\in
\arg\max_{v\in V\setminus N_j}\eta_j(v),
\;
N_{j+1}=N_j\cup\{v^*\},
\;
\mu_{j+1}(q_\mathsf{new})=v^*,
$$
where $q_\mathsf{new}$ is the newly adopted data patch initialized in $|+\rangle$ \cite{van2020zx}, thereby exposing as many blocked interactions as possible.
Alg. \ref{alg:execution-order} summarizes these transitions.

\begin{algorithm}[t]
\caption{Execution ordering of a reduced ZX-diagram}
\label{alg:execution-order}
\begin{spacing}{1}
\begin{algorithmic}[1]
\footnotesize
\Require Reduced ZX-diagram $D_r=(V,E,\phi)$
\Ensure Execution sequence $\pi$

\State $N\gets V_{\mathrm{in}}$, $R\gets E$, $\pi\gets[\,]$

\While{$R\neq\emptyset$}

    \While{$\mathcal X(N,R)\neq\emptyset$}
        \State $L\gets
        \mathsf{MaximalMatching}(\mathcal X(N,R))$
        \State emit $\mathsf{CZ}(L)$ into $\pi$; $R\gets R\setminus L$
    \EndWhile

    \State $\mathcal F\gets
    \{u\in N\setminus V_{\mathrm{out}}\mid\deg_R(u)=0\}$
    \State $N\gets N\setminus \mathcal F$

    \State $\mathcal H\gets
    \mathsf{DisjointMoves}(\mathcal H(N,R))$
    \If{$\mathcal H\neq\emptyset$}
        \State emit $\mathsf{H}(\mathcal H)$ and update $N$, $\mu$, and $R$
        \State \textbf{continue}
    \EndIf

    \If{$R\neq\emptyset$ and $\mathcal X(N,R)=\emptyset$}
        \State
        $v^*\gets\arg\max_{v\notin N}\eta(v)$
        \State activate $v^*$ in $N$ and update $\mu$
    \EndIf
\EndWhile

\State \Return $\pi$
\end{algorithmic}
\end{spacing}
\end{algorithm}

\begin{figure}[ht]
     \centering
     \begin{subfigure}{0.23\textwidth}
         \centering
         \includegraphics[width=0.83\textwidth]{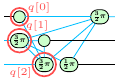}
         \caption{Step 1. Three data patches are initialized, and two CZ operations are extracted.}
         \label{fig:sample_zx_order1}
     \end{subfigure}
     \hfill
     \begin{subfigure}{0.23\textwidth}
         \centering
         \includegraphics[width=0.83\textwidth]{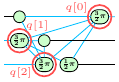}
         \caption{Step 2. $q[0]$ moves from $v[0]$ to $v[3]$, and two CZ operations are extracted.}
         \label{fig:sample_zx_order2}
     \end{subfigure}
     \\
     \begin{subfigure}{0.23\textwidth}
         \centering
         \includegraphics[width=0.83\textwidth]{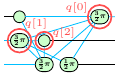}
         \caption{Step 3. $q[2]$ moves from $v[2]$ to $v[4]$, and one CZ operation is extracted.}
         \label{fig:sample_zx_order3}
     \end{subfigure}
     \hfill
     \begin{subfigure}{0.23\textwidth}
         \centering
         \includegraphics[width=0.83\textwidth]{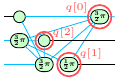}
         \caption{Step 4. $q[1]$ moves from $v[1]$ to $v[5]$, and one CZ operation is extracted, ending the process.}
         \label{fig:sample_zx_order4}
     \end{subfigure}
     \caption{Extracting the execution order from the full-reduced ZX-diagram. $H$ operations on the input and output edges are neglected.}
     \label{fig:sample_zx_order}
     \Description{}
\end{figure}

Fig. \ref{fig:sample_zx_order} applies Algorithm \ref{alg:execution-order} to the fully reduced graph in Fig. \ref{fig:sample_zx_reduce}.
The three input data patches initially occupy:
$$
N_0=\{v[0],v[1],v[2]\},
$$
which makes the edges $\{v[0],v[2]\}$ and $\{v[1],v[2]\}$ immediately executable.
After these CZ interactions are consumed, the active frontier evolves through
\begin{align*}
\{v[0],v[1],v[2]\}
\xrightarrow{\mathsf H_{0\rightarrow3}}
\{v[3],v[1],v[2]\}\\
\xrightarrow{\mathsf H_{2\rightarrow4}}
\{v[3],v[1],v[4]\}
\xrightarrow{\mathsf H_{1\rightarrow5}}
\{v[3],v[5],v[4]\}.
\end{align*}
Each frontier exposes the CZ interactions.
Note that input and output boundary H edges are omitted, since they correspond to boundary type switches before and after the internal computation \cite{litinski2018lattice}.

\begin{figure}[ht]
     \centering
     \includegraphics[width=0.4\textwidth]{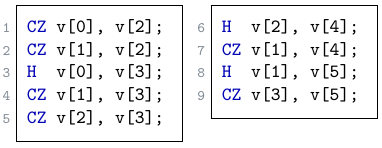}
     \caption{Execution order of the internal edges. $H$ operations mark the boundaries of each time step.}
     \label{fig:sample_order}
     \Description{}
\end{figure}

The resulting sequence $\pi$, including the three H transitions that separate successive frontiers, is shown in Fig. \ref{fig:sample_order}.

\subsection{Target Code Generation}

The previous step produces an execution sequence $\pi$ together with the phase of each reduced ZX spider. 
SpiderLS reconstructs a logical target program as 
$$
\langle \pi,\phi\rangle \xrightarrow{\mathsf{group}} \overline G_{\mathrm{tgt}}. 
$$

Since the input is restricted to Clifford+T circuits, every spider phase is an integer multiple of $\pi/4$:
$$
\phi(v)=\frac{\kappa_v\pi}{4}, \qquad \kappa_v\in\mathbb Z_8 . 
$$
The seven nonzero cases are grouped and handled as follows:
\begin{enumerate}
    \item $\kappa_v = 4$: The phase corresponds to a $Z$ gate, which can be tracked in the Pauli frame and therefore requires no physical execution \cite{PennyLane-PauliFrameTracking, riesebos2017pauli}.
    \item $\kappa_v = 2$: The phase corresponds to the $\sqrt{Z}=S$ gate. 
    By applying the inverse of the Fusion rule in Fig. \ref{fig:zx_rule}, the phase spider can be unfused into the $Y$ basis measurement spider described in Sec. \ref{sec:logical_patches}. 
    In the spacetime diagram, this primitive is represented by a green block.
    \item $\kappa_v = 6$: The phase corresponds to $S^\dagger=ZS$, and can therefore be reduced to $S$ using Pauli frame.
    \item $\kappa_v = 1, 7$: These phases correspond to $\sqrt{S}=T$ and $T^\dagger$, respectively. 
    They are implemented through T gate injection from a factory located at the boundary of the patch grid, using a $ZZ$ measurement. 
    The injection yields $T$ and $T^\dagger$ with equal probability, and an undesired outcome can be \emph{corrected} using a conditional $Y$ basis measurement \cite{akahoshi2024partially,sethi2025rescq}. 
    The correction can also be performed externally using an additional ancilla, allowing the injection to take a single time step \cite{zhou2026topols}. 
    Therefore, $T$ and $T^\dagger$ can be treated as equivalent operations, and $T^\dagger$ is represented as $T$ in the target code.
    The corresponding injection port is represented by a purple block in the spacetime diagram.
    \item $\kappa_v = 3, 5$: These phases correspond to $ZT^\dagger$ and $ZT$, and are reduced to $T^\dagger$ and $T$ using Pauli frame.
\end{enumerate}
Consequently, the target code exposes only $S$ and $T$ phase operations while preserving the exact phase semantics:
$$
\Phi (v) = \begin{cases} I, & \kappa_v\in\{0,4\},\\ S, & \kappa_v\in\{2,6\},\\ T, & \kappa_v\in\{1,3,5,7\}. \end{cases}
$$
A phase operation associated with spider $v$ is emitted when the data patch occupying $v$ completes its interactions at that spider. 
For a patch $q$ with $\mu(q)=v$, 
$$
\Phi(v)=S \Rightarrow S(q), \qquad \Phi(v)=T \Rightarrow T(q). 
$$

SpiderLS next exploits the higher-arity interactions motivated in Sec. \ref{sec:motivation2}. 
Within each execution step, CZ interactions sharing the same logical control are greedily combined into a maximal multi-target CZ: 
$$
\mathsf{CZ}(q_c,q_1); \cdots; \mathsf{CZ}(q_c,q_k) \;\Longrightarrow\; \mathsf{CZ}^{(k)}(q_c;q_1,\ldots,q_k). 
$$
For $k=1$, $\mathsf{CZ}^{(1)}$ is an ordinary CZ. 
Thus, the target language is defined as
$$
g ::= \mathsf H(q) \mid \mathsf S(q) \mid \mathsf T(q) \mid \mathsf{CZ}^{(k)}(q_c;q_1,\ldots,q_k),
$$
$$
\overline G_{\mathrm{tgt}} ::= g_1;\cdots;g_r . 
$$

\begin{figure}[ht]
     \centering
     \includegraphics[width=0.45\textwidth]{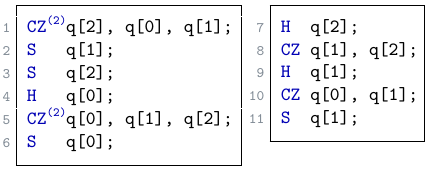}
     \caption{Final target code with phase operations. Two CZ operations merge into a double-target CZ operation, where the first qubit is the control qubit. $\mathsf{CZ}^{(1)}$ is denoted as $\mathsf{CZ}$.}
     \label{fig:sample_target}
     \Description{}
\end{figure}

Fig. \ref{fig:sample_target} shows the resulting target code for the running example. 
The first two CZ interactions in the execution order share $q[2]$ and are combined as 
$$
\mathsf{CZ}^{(2)}(q[2];q[0],q[1]). 
$$
After $q[0]$ advances to the next spider, the next two interactions are similarly combined into 
$$
\mathsf{CZ}^{(2)}(q[0];q[1],q[2]). 
$$
The remaining interactions are emitted as ordinary CZ gates, while the phases attached to the spiders produce the $S$ gates.

\subsection{PPM Lowering and Logical Scheduling}

SpiderLS next lowers the target program to a sequence of unit-time PPMs, and then packs the resulting measurements into logical layers.
We write
$$
\overline G_{\mathrm{tgt}}
\Downarrow_{\mathrm{PPM}}
\overline M,
\qquad
\overline M = m_1;\cdots;m_s .
$$

Each target instruction is lowered to PPMs as follows:
$$
\begin{aligned}
\mathsf{CZ}^{(k)}\!(q_c;q_1,\!\ldots\!,q_k)\!
&\Downarrow_{\mathrm{PPM}}
M_{Z^kX}\!(q_1,\!\ldots\!,q_k,q_A);
M_{ZZ}\!(q_c,q_A),\\
\mathsf S(q)
&\Downarrow_{\mathrm{PPM}}
M_{ZZ}(q,q_A);
M_Y(q_A),\\
\mathsf T(q)
&\Downarrow_{\mathrm{PPM}}
M_{ZZ}(q,q_{\mathbf{magic}}),\\
\mathsf H(q)
&\Downarrow_{\mathrm{PPM}}
\delta_H(q).
\end{aligned}
$$
Here, $q_A$ denotes an internal ancilla qubit and $q_{\mathbf{magic}}$ an externally supplied magic state resource.
The Hadamard event $\delta_H(q)$ is treated as a zero-time boundary-frame update that logically swaps the $X$  and $Z$ boundary interpretations \cite{litinski2018lattice}.

The lowered PPMs form a DAG whose edges encode measurement dependencies, including the ordering between stages of the same target operations and per-patch dependencies across $H$ operations and measurements/initializations.
SpiderLS greedily packs ready PPMs into logical layers while respecting the available ancilla qubit patches and the boundary capacity of each data qubit patch.  
Under the square patch model, at most two physical $X$ and two $Z$ boundaries of a patch may be used in the same layer.

\begin{figure}[t]
     \centering
     \includegraphics[width=0.48\textwidth]{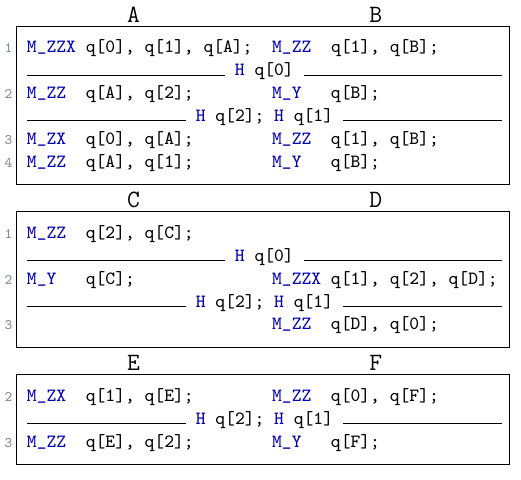}
     \caption{The PPM sequence is maximally parallelized into four logical layers using three data patches and six internal ancilla patches.
     Columns A--F denote persistent ancilla patches, and the stages of the same gadget remain in the same column.
     $H$ operations are zero-time events inserted between logical layers and do not contribute additional depth.}
     \label{fig:sample_parallel}
     \Description{}
\end{figure}

To expose maximal parallelism under this scheduler while avoiding unnecessary ancilla patches, SpiderLS first computes an unbounded ancilla schedule and uses its depth as a baseline:
$$
d^\star =
\operatorname{depth}
\bigl(\mathsf{Sched}(\overline M,\infty)\bigr),
$$
where $\infty$ means that an infinite number of ancillae is used.
It then increases the ancilla limit from one and selects the smallest limit that reproduces the same depth:
$$
a^\star =
\min\left\{
a \;\middle|\;
\operatorname{depth}
\left(\mathsf{Sched}(\overline M,a)\right)=d^\star
\right\}.
$$
The resulting logical schedule is therefore
$$
\Sigma=\mathsf{Sched}(\overline M,a^\star),
$$
which preserves the depth of the unbounded ancilla schedule using the fewest internal ancillae found by the search.

Fig. \ref{fig:sample_parallel} shows the schedule obtained from the target code of Fig. \ref{fig:sample_target}.
The resulting PPMs are packed into four logical layers.
Columns A--F correspond to internal ancilla patches $q_A,\ldots,q_F$, and the two stages of each gadget remain in the same column.
The $H$ operations inserted between consecutive layers are zero-time events and therefore do not increase the number of logical time steps.
The schedule uses three data patches and six persistent ancilla patches, and serves as the input to the routing phase described next.

\begin{algorithm}[t]
\caption{Layered spacetime routing}
\label{alg:routing}
\begin{spacing}{1}
\begin{algorithmic}[1]
\footnotesize
\Require Logical PPM schedule $\Sigma$
\Ensure Spacetime realization $S$
\State Fix data qubit positions
\For{physical layer $t=1,2,\ldots$}
    \State $\overline M_R\gets$ dependency ready PPMs with source layer $\le t$
    \For{$m\in \overline M_R$}
        \State place a new ancilla, if required
        \State $\rho\gets\Call{Route}{m}$
        \If{$\rho$ succeeds}
            \State commit $\rho$ to layer $t$
        \Else
            \State defer $m$
        \EndIf
    \EndFor
    \State apply ready $H$ operations
    \If{all PPMs are routed}
        \State \Return spacetime realization $S$
    \EndIf
\EndFor
\end{algorithmic}
\end{spacing}
\end{algorithm}

\subsection{Layered Spacetime Routing}

SpiderLS maps the logical PPM schedule $\Sigma$ to a lattice surgery spacetime realization:
$$
\Sigma \xrightarrow{\mathsf{route}} S .
$$
The routing summarized in Alg. \ref{alg:routing} proceeds as follows.
\begin{enumerate}
    \item \textbf{Data qubit placement.} Data qubit patches are placed once in a near-square snake layout and remain fixed throughout the computation.
    
    \item \textbf{Measurement ordering.}
    At the beginning of each physical layer, SpiderLS freezes the set of dependency-ready PPMs whose logical source layer has been reached, and attempts them in the following priority order:
    \begin{enumerate}
        \item earlier logical source layer;
        \item measurements that complete their gate operations;
        \item larger number of terminals;
        \item larger spatial span among participating data qubits.
    \end{enumerate}

    \item \textbf{Ancilla placement.}
    If a PPM introduces a new internal ancilla, candidate positions and orientations are considered in the following priority order:
    \begin{enumerate}
        \item smaller summed Manhattan distance to the required data
        qubit contacts;
        \item smaller expansion of the data patch bounding box;
        \item smaller distance to the center of the data layout.
    \end{enumerate}
    Once placed, the ancilla position and orientation remain fixed until the final measurement of the gate.

    \item \textbf{PPM routing.}
    For each PPM, SpiderLS searches a bounded set of compatible
    terminal-side assignments determined by the required Pauli
    boundaries. Connections are constructed using exact shortest-path
    A* \cite{hart1968formal}; for multi-terminal PPMs, a rectilinear Steiner tree is grown
    greedily toward the nearest remaining terminal using A* for each
    extension \cite{hwang1992steiner}.
    For $M_{ZZ}(q,q_{\mathrm{magic}})$, the magic state is supplied
    externally, so its route terminates at the boundary of patch
    grid.
    Routes within the same physical layer are disjoint.

    \item \textbf{Commit and deferral.}
    A successfully routed PPM is immediately committed to the current physical layer. 
    If no valid route is found under the bounded local search, the PPM is deferred to a later layer, while other ready PPMs may still use the remaining routing space. Ready $H$ operations are applied after the successful measurements of the layer are committed.
\end{enumerate}

\begin{figure}[b]
     \centering
     \begin{subfigure}{0.15\textwidth}
         \centering
         \includegraphics[width=\textwidth]{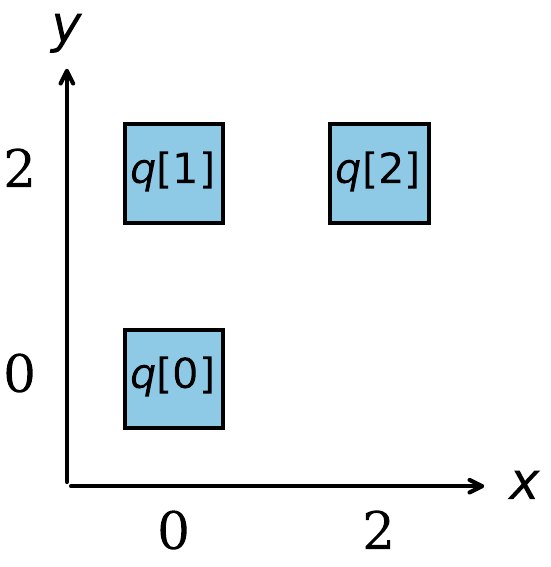}
         \caption{Data qubits are initialized at $(0,0)$, $(0,2)$, and $(2,2)$.}
         \label{fig:sample_routing1}
     \end{subfigure}
     \hfill
     \begin{subfigure}{0.15\textwidth}
         \centering
         \includegraphics[width=\textwidth]{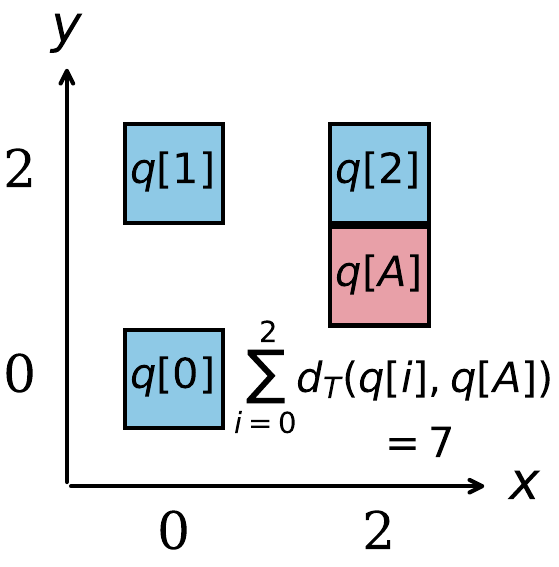}
         \caption{Ancilla qubit $q[A]$ is initialized at $(2,1)$.}
         \label{fig:sample_routing2}
     \end{subfigure}
     \hfill
     \begin{subfigure}{0.15\textwidth}
         \centering
         \includegraphics[width=\textwidth]{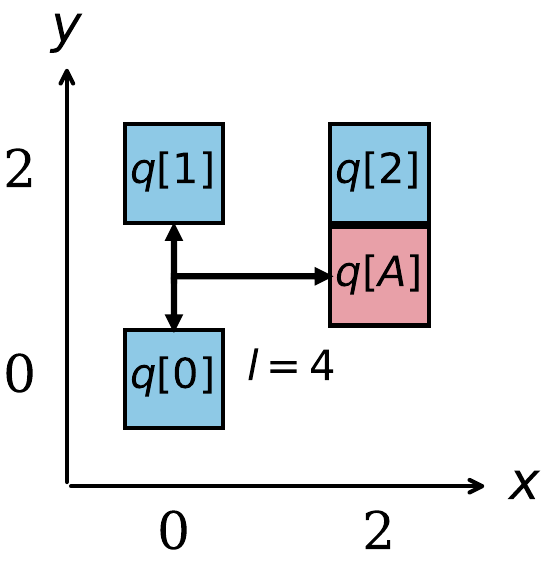}
         \caption{The PPM connection route is determined.}
         \label{fig:sample_routing3}
     \end{subfigure}
     \caption{Initialization of the data qubit patches and routing of the first measurement operation, $M_{ZZX}(q_0,q_1,q_A)$, which is placed at $(2,1)$, where the sum of Manhattan distances ($d_T$) is 7. The length of the PPM connection route is 4.}
     \label{fig:sample_routing}
     \Description{}
\end{figure}

\begin{figure}[t]
     \centering
     \begin{subfigure}{0.27\textwidth}
         \centering
         \includegraphics[width=\textwidth]{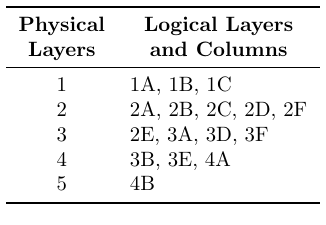}
         \caption{Five physical layers are needed to route the measurement operations.}
         \label{fig:sample_pl}
     \end{subfigure}
     \hfill
     \begin{subfigure}{0.19\textwidth}
         \centering
         \includegraphics[width=\textwidth]{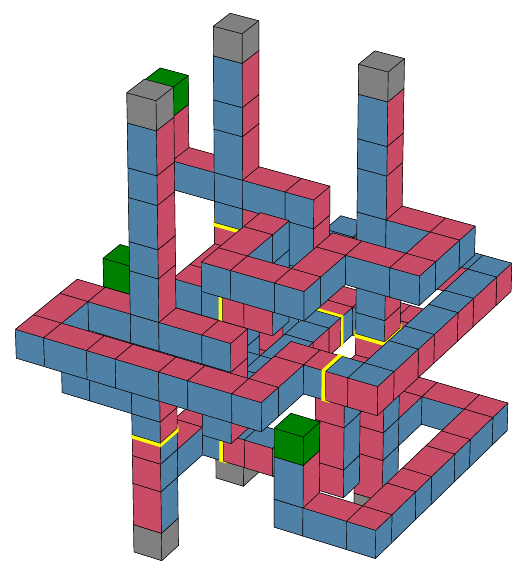}
         \caption{The spacetime diagram of the routed result.}
         \label{fig:sample_diagram}
     \end{subfigure}
     \caption{Measurement operations are greedily routed on the physical layers, after which the corresponding spacetime diagram can be constructed. The green blocks represent $Y$ basis measurements.
     The spacetime volume is $25 \times 5=125$.}
     \label{fig:sample_spacetime}
     \Description{}
\end{figure}

For the running example, Fig. \ref{fig:sample_routing} shows the placement and routing of the first PPM, $M_{ZZX}(q[0],q[1],q[A])$.
After fixing the three data patches, $q[A]$ is placed at $(2,1)$, where the summed Manhattan distance is 7, and the resulting connection has route length 4.
Fig. \ref{fig:sample_spacetime} then shows the complete result: the four logical layers of Fig. \ref{fig:sample_parallel} expand to five physical layers due to routing conflicts, and stacking these layers produces the final spacetime diagram.

\section{Evaluation}

We evaluate SpiderLS in terms of spacetime cost, compilation time, and magic state demand across structured and random benchmarks.

\subsection{Experimental Setup}

\paragraph{Baselines.}
We compare SpiderLS against three scalable lattice surgery compilation frameworks: Liblsqecc \cite{watkins2024high}, DASCOT \cite{molavi2025dependency}, and TopoLS \cite{zhou2026topols}.
Liblsqecc is a circuit-based compiler that prioritizes low compilation overhead and scalability over aggressive spacetime optimization.
DASCOT follows the circuit-based approach but introduces dependency-aware optimization to expose additional parallelism and reduce the number of execution time steps.
TopoLS instead operates on a ZX-diagram representation, using spider fusion and layer slicing followed by MCTS-based spacetime embedding.
These baselines therefore represent progressively stronger optimization strategies.

\paragraph{Benchmarks.}
We evaluate SpiderLS on a diverse set of algorithmic and random circuits.  Bernstein--Vazirani (BV) and Deutsch--Jozsa (DJ) are evaluated at 16, 20, 40, 60, 80, and 100 qubits, while Grover search uses 6 qubits.  
GHZ, QAOA, QFT, QPE, VQE, and W-state circuits are evaluated at 16 qubits.  We additionally use random Clifford circuits with 4, 6, 8, 10, 12, 16, 20, 40, and 60 qubits to study scalability independently of a specific algorithmic structure.  
To evaluate magic state demand, we construct 6-qubit random Clifford+T circuits by randomly inserting T gates into the Clifford circuits with insertion probabilities $p_T\in\{0.2,0.4,0.6,0.8\}$.

\paragraph{Metrics.}
Our primary metric is the spacetime volume, which captures the combined spatial and temporal cost of the generated lattice surgery computation.
We additionally report compilation time to evaluate compiler scalability.

For circuits containing T gates, we measure \emph{T-port density} to quantify the concentration of magic state demand on the spacetime boundary.
Let $N_T$ denote the number of external T injection ports and let $W$, $L$, and $H$ denote the width, length, and height of the bounding spacetime cuboid.
We define
$
\rho_T
=
N_T/\left(2(W+L)H\right),
$
where $2(W+L)H$ is the total area of the four lateral faces of the cuboid.
A larger $\rho_T$ therefore indicates that magic states must be supplied more densely per unit lateral spacetime area.

\paragraph{Platform.}
All experiments are performed on an x86 system with an Intel\textsuperscript{\textregistered} Core\textsuperscript{TM} i7-11800H CPU at 2.30\,GHz \cite{intel_i7_11800h_ark} and 48.0\,GB of DDR4 memory.
This commodity laptop-class platform allows us to evaluate whether the compilation workflow remains practical without relying on server-class computing resources.

\begin{figure}[t]
     \centering
     \begin{subfigure}{0.49\textwidth}
         \centering
         \includegraphics[width=\textwidth]{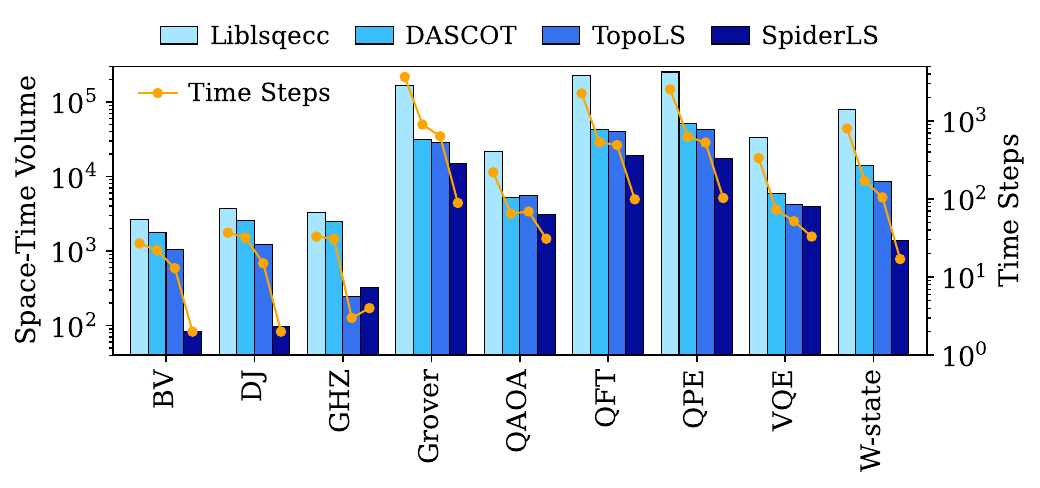}
         \caption{Spacetime volume and time step comparison.}
         \label{fig:spacetime_volume}
     \end{subfigure}
     \\
     \begin{subfigure}{0.46\textwidth}
         \centering
         \includegraphics[width=\textwidth]{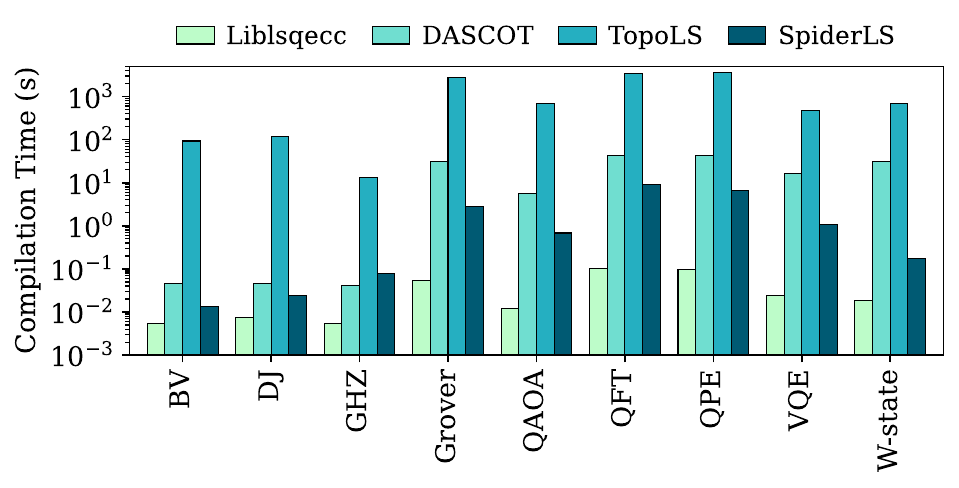}
         \caption{Compilation time comparison.}
         \label{fig:compilation_time}
     \end{subfigure}
     \caption{Overall comparison of spacetime volume and compilation time of Liblsqecc, DASCOT, TopoLS, and SpiderLS.
     All vertical axes are logarithmic.}
     \label{fig:experiment_overall}
     \Description{}
\end{figure}

\subsection{Overall Comparison}

Fig. \ref{fig:experiment_overall} compares the spacetime volume, number of time steps, and compilation time of the four compilers.
All vertical axes are shown on logarithmic scales.
On average, SpiderLS reduces spacetime volume and compilation time by 49.2\% and 99.8\% compared with the SOTA framework, TopoLS.

Fig. \ref{fig:spacetime_volume} shows that the spacetime volume generally decreases from Liblsqecc to DASCOT, TopoLS, and SpiderLS, with the corresponding time-step results exhibiting the same overall trend.  
This indicates that temporal reduction is the main contributor to the lower spacetime volume.  
In particular, SpiderLS exploits the multi-patch measurement strategy described earlier by grouping CZ interactions with a common control into a single multi-patch measurement.
GHZ is an illustrative exception because its interaction structure provides little opportunity for such grouping.  
Most other benchmarks expose more opportunities for multi-patch measurements and therefore achieve substantial reductions in both time steps and spacetime volume.

Fig. \ref{fig:compilation_time} shows that SpiderLS compiles all evaluated benchmarks substantially faster than TopoLS and is also faster than DASCOT in most cases, while Liblsqecc generally retains the smallest compilation overhead due to its lightweight optimization strategy.   
This supports the structure-aware routing approach of SpiderLS: rather than using MCTS to search coupled embedding choices within a layer, SpiderLS performs bounded local routing on already-lowered PPMs.

\begin{figure}[t]
     \centering
     \includegraphics[width=0.48\textwidth]{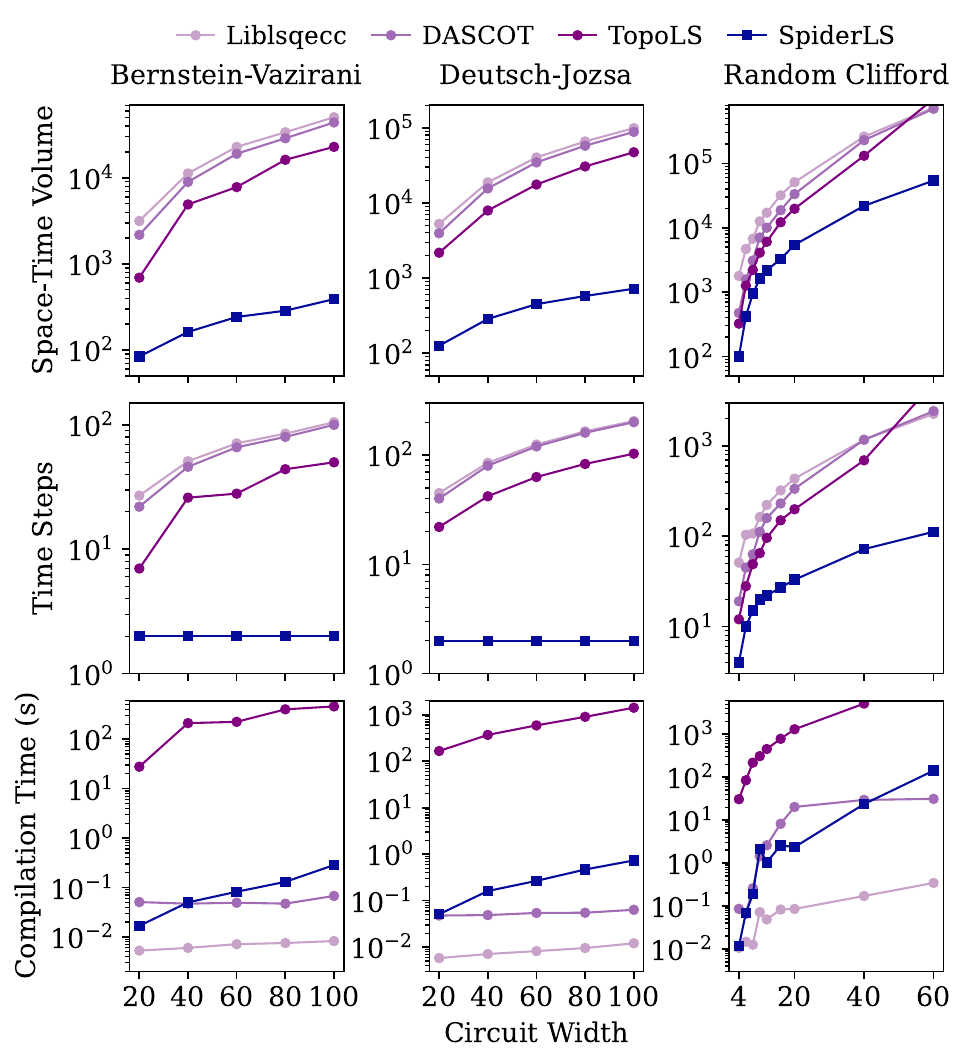}
     \caption{Scalability comparison with increasing circuit width for BV, DJ, and random Clifford circuits.}
     \label{fig:scalability}
     \Description{}
\end{figure}

\subsection{Scalability Evaluation}

Fig. \ref{fig:scalability} evaluates scalability with increasing circuit width for BV, DJ, and random Clifford circuits.
All vertical axes use logarithmic scales.

SpiderLS consistently produces substantially smaller spacetime volumes as the circuit width increases.
The reduction is accompanied by nearly constant time step counts for BV and DJ and a much slower growth for random Clifford circuits, showing that the advantage is preserved at larger problem sizes.
Despite this aggressive spacetime reduction, the compilation time of SpiderLS remains close to Liblsqecc and DASCOT over most configurations, while staying orders of magnitude below TopoLS.
These results show that SpiderLS improves spacetime efficiency without incurring the large compilation overhead associated with search-based spacetime embedding.

\begin{figure}[t]
     \centering
     \includegraphics[width=0.4\textwidth]{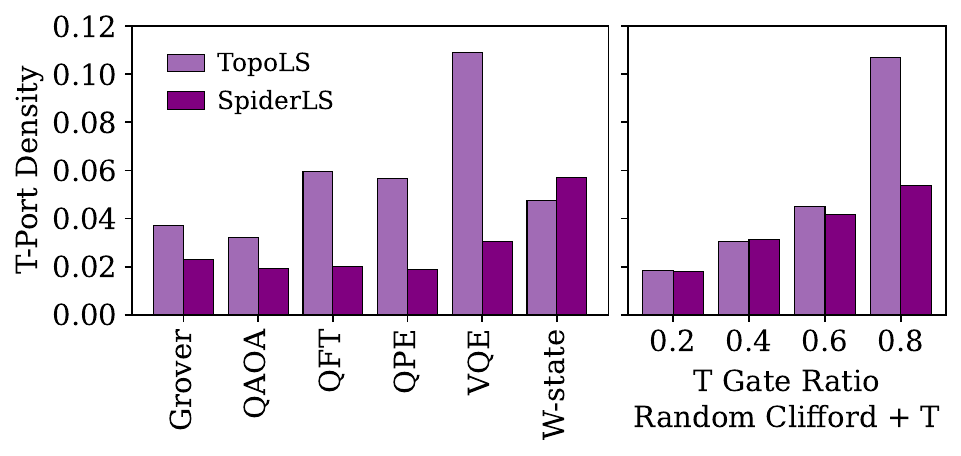}
     \caption{T-port density comparison between TopoLS and SpiderLS for algorithmic and random Clifford+T circuits.}
     \label{fig:t_density}
     \Description{}
\end{figure}

\subsection{Magic State Demand}

Fig. \ref{fig:t_density} compares the T-port density of TopoLS and SpiderLS for the benchmarks that contain T gates.
Despite using fewer time steps, SpiderLS maintains a T-port density comparable to or lower than TopoLS for most benchmarks.
Since reducing temporal depth decreases the lateral area available for T injection, this result indicates that the temporal compression does not come at the cost of more concentrated magic state demand.
This behavior is consistent with the effect of ZX reduction, which simplifies and merges phase structure and thereby reduces redundant T injections.
The effect is also preserved as the T insertion probability increases in the random Clifford+T benchmarks.

\section{Related Work}

Existing scalable lattice surgery compilers are largely circuit-oriented.
Liblsqecc \cite{watkins2024high} emphasizes fast compilation with lightweight optimization, while DASCOT \cite{molavi2025dependency} further exploits circuit dependencies to reduce execution depth.
Related routing approaches such as edge-disjoint-path compilation \cite{beverland2022surface} also improve the parallel realization of long-range surface code operations.
Recent work on multi-qubit lattice surgery scheduling \cite{silva_et_al:LIPIcs.TQC.2024.1} goes beyond pairwise gates by representing multi-qubit operations with routing trees and scheduling them directly.
SpiderLS is complementary to these approaches in that it first exposes multi-patch interactions through ZX reduction and ordering before performing logical scheduling and spacetime routing.

Solver-based approaches \cite{liao2026design} such as LaSsynth \cite{tan2024sat} target a different use case.
They formulate lattice surgery synthesis as a SAT problem and searches for volume-optimal realizations of small subroutines.
Since their primary goal is subroutine-level synthesis rather than scalable end-to-end compilation, we do not include them as a baseline.

ZX-calculus provides a natural representation for lattice surgery \cite{de2020zx}.
Topologiq \cite{Bolanos_Topologiq_Algorithmic_Lattice_2025} uses ZX-diagram connectivity to incrementally construct topologically valid lattice surgery pipe diagrams, but primarily serves as a geometric construction engine rather than an end-to-end optimization framework.
TopoLS \cite{zhou2026topols} extends this direction with ZX simplification, layer slicing, and MCTS-based spacetime embedding.

\section{Conclusion}

We present SpiderLS, a lattice surgery compiler that exploits multi-patch measurements and structure-aware routing.
SpiderLS substantially reduces spacetime volume while maintaining low compilation overhead, demonstrating the benefits of full ZX reduction for lattice surgery compilation.

\section*{Data Availability}

The SpiderLS implementation, benchmark circuits, and reproduction scripts are
publicly available at \url{https://github.com/fluorite42/SpiderLS}.





\bibliographystyle{ACM-Reference-Format}
\bibliography{acmart}

\end{document}